\documentclass[graphics,floatfix, footinbib,tightenlines,nobibnotes, aps, prl, twocolumn]{revtex4-1}
\usepackage{makecell}
\usepackage{amsmath,amssymb,wasysym}
\usepackage{graphicx}
\usepackage{dcolumn}
\usepackage{bm}
\usepackage{braket}
\usepackage{verbatim}
\usepackage{float}
\usepackage{bbold}
\usepackage{color}
\usepackage{xcolor}
\usepackage{relsize}
\usepackage{amsthm}
\usepackage{enumerate}
\usepackage{soul,xcolor}
\usepackage[T1]{fontenc}
\usepackage{adjustbox}
\usepackage{hyperref}
\hypersetup{
colorlinks = true,
linkcolor = [rgb]{0.70,0.13,0.13},
citecolor = [rgb]{0.13,0.55,0.13},
urlcolor = [rgb]{0.25, 0.41, 0.88}}

\newcommand{\be}{\begin{equation}}
\newcommand{\ba}{\begin{align}}
\newcommand{\ee}{\end{equation}}
\newcommand{\bea}{\begin{eqnarray}}
\newcommand{\eea}{\end{eqnarray}}
\newcommand{\beq}{\begin{equation}}
\newcommand{\eeq}{\end{equation}}
\newcommand{\beqn}{\begin{eqnarray}}
\newcommand{\eeqn}{\end{eqnarray}}

\renewcommand{\hat}[1]{{\widehat #1}}

\definecolor{Red}{rgb}{0.70,0.13,0.13}
\usepackage{bm}

\usepackage{array}
\newcolumntype{L}[1]{>{\raggedright\arraybackslash}p{#1}}
\newcolumntype{C}[1]{>{\centering\arraybackslash}p{#1}}
\newcolumntype{R}[1]{>{\raggedleft\arraybackslash}p{#1}}
\usepackage{multirow}
\allowdisplaybreaks

\begin{document}

\title{Type II t-J model in charge transfer regime in bilayer La$_3$Ni$_2$O$_7$ and trilayer La$_4$Ni$_3$O$_{10}$}

\author{Hanbit Oh }
\thanks{These two authors contributed equally}
\email{hoh22@jh.edu}
\author{Boran Zhou }
\thanks{These two authors contributed equally}

\author{Ya-Hui Zhang}
\email{yzhan566@jhu.edu}

\affiliation{William H. Miller III Department of Physics and Astronomy, Johns Hopkins University, Baltimore, Maryland, 21218, USA}
\begin{abstract}
Recent observations of an 80 K superconductor in La$_3$Ni$_2$O$_7$ under high pressure have attracted significant attention.   Recent experiments indicate that La$_3$Ni$_2$O$_7$ may be in the charge transfer regime, challenging the previous models based purely on the Ni $d_{x^2-y^2}$ and $d_{z^2}$ orbitals. In this study, we propose a low energy model that incorporates doped holes in the oxygen $p$ orbitals. Given that the parent nickel state is in the $3d^{8}$ configuration with a spin-one moment, doped hole only screens it down to spin-half, in contrast to the Zhang-Rice singlet in cuprate. We dub the single hole state as Zhang-Rice spin-half and build an effective model which includes three spin-one states ($d^8$) and two Zhang-Rice spin-half states ($d^8 L$). At moderate pressure around $20$ GPa, the dominated oxygen orbital is an in-plane Wannier orbital with the same lattice symmetry as the $d_{x^2-y^2}$ orbital. 
The resulting model reduces to the bilayer type II t-J model previously proposed in the Mott-Hubbard regime. Notably, the hopping between the in-plane $p$ orbitals of the two layers is still suppressed. Density matrix renormalization group (DMRG) simulation reveals a pairing dome with the optimal hole doping level at $x=0.4\sim0.5$, distinct from the hole doped cuprate where optimal doping occurs around $x=0.19$. Further increasing pressure initially raises the critical temperature ($T_c$) until reaching an optimal pressure beyond which the $p_z$ orbital of oxygen becomes favorable and superconductivity is diminished. This shift from in-plane $p$ orbital to $p_z$ orbital may elucidate the experimentally observed superconducting dome with varying pressure.  As an extension, we also suggest a trilayer version of the type II t-J model as the minimal model for pressured La$_4$Ni$_3$O$_{10}$, which is distinct from the models in the Mott-Hubbard regime.
 \end{abstract}

\maketitle

{\it Introduction.}--- The recent discovery of a superconductor with a critical temperature of approximately 80 K in La$_3$Ni$_2$O$_7$ under high pressure\cite{sun2023signatures,hou2023emergence,zhang2023hightemperature} has sparked considerable interest\cite{liu2023electronic,yang2023orbital,zhang2023effects, PhysRevLett.131.126001,Zhang_2023,yang2023possible,sakakibara2023possible,gu2023effective,shen2023effective,wu2023charge,PhysRevLett.131.206501,liu2023s,cao2023flat,PhysRevB.108.174511,PhysRevLett.132.146002,qu2023bilayer,lu2023superconductivity,jiang2023pressure,tian2023correlation,zhang2023strong,qin2023high,huang2023impurity,zhang2023trends,jiang2023high,yang2023minimal,2023arXiv230809044Q,2023arXiv230807386Z,2023arXiv230812750K,2023arXiv230811614J,fabian1,fabian2,fabian3,jiang2024pressure,zhan2024cooperation,lechermann2024electronic,wang2024pressure,xue2024magnetism,kaneko2024pair,jiang2024high,ryee2023critical,PhysRevB.108.214522,wu2023charge}. Additionally, there is emerging evidence of superconductivity in  La$_4$Ni$_3$O$_{10}$ under high pressure, exhibiting critical temperatures ranging from 20-30 K, which has further fueled research interest in this area \cite{Li__2024,zhu2024superconductivity,sakakibara2024theoretical,zhang2024superconductivity,lu2024superconductivity,zhang2024prediction,labollita2024electronic,yang2024effective}. Identifying a minimal model that captures the essential physics is a crucial step forward.

According to density functional theory (DFT) \cite{sun2023signatures}, the valence of the nickel (Ni) atom in the bilayer nickelate La$_3$Ni$_2$O$_7$ is in the $3d^{8-x}$ configuration.  The $d_{x^2-y^2}$ orbital is at density  $n_1=1-x$ per site with $x=0.5$, while the $d_{z^2}$ orbital has the density $n_2\approx 1$ and is near Mott localization. 
Following this picture, many theoretical works propose a two-orbital Hubbard or t-J model in terms of the $d_{x^2-y^2}$ and the $d_{z^2}$ orbital. Especially, Refs.\cite{PhysRevB.108.174511,PhysRevLett.132.146002} highlight the role of Hund's coupling $J_H$ in transmitting a large inter-layer spin-spin coupling $J_\perp$ from the $d_{z^2}$ to the $d_{x^2-y^2}$ orbital, providing a plausible explanation for a high critical temperature ($T_c$) even at large doping $x\approx 0.5$.
In Ref.\cite{PhysRevLett.132.146002}, the localized $d_{z^2}$ orbitals are simply integrated out, so the model is reduced to a one-orbital bilayer t-J model of $d_{x^2-y^2}$ with a negligible $t_{\perp}$. But the integration of $d_{z^2}$ orbital is not appropriate in the large $J_H$ limit \cite{PhysRevB.108.174511,yang2023strong}. Instead, the minimal model is demonstrated to be a bilayer type-II t-J model \cite{PhysRevB.108.174511,yang2023strong}, which includes the spin-one Ni$^{2+}$ state (3$d^8$) and spin-half Ni$^{3+}$ state (3$d^7$) at each site. The type II t-J model hosts unique physics such as a dome of pairing gap around $x\approx 0.5$ due to doping induced Bardeen-Cooper-Schrieffer(BCS) to Bose-Einstein-condensate (BEC) crossover\cite{yang2023strong}. There are also two different Fermi liquids above $T_c$ with a jump of the Fermi surface volume by $1/2$ Brillouin zone\cite{PhysRevB.108.174511,yang2023strong,wu2024deconfined}. Clearly the physics is essentially different from the familiar hole doped cuprates. However, the model is derived\cite{PhysRevB.108.174511} assuming  that the system is in the Mott-Hubbard regime of the 
 Zaanen–Sawatzky–Allen classification scheme \cite{PhysRevLett.55.418}. 
Depending on the energy splitting $\Delta$ between the oxygen $p$ orbital and the Ni $d$ orbital compared to Hubbard U, we have the Mott-Hubbard regime ($\Delta>U$) or the charge-transfer regime ($\Delta<U$).
In the Mott-Hubbard regime, the doped holes enter the 3d orbitals of the Ni atom and we can ignore the oxygens. 
However, a recent experiment suggests that  La$_3$Ni$_2$O$_7$ might be within the charge transfer regime 
where holes enter the oxygen $p$ orbitals while the nickel atom is pinned to be in the 3$d^8$ configuration with a localized spin-one moment \cite{dong2023visualization}. This challenges the previous theoretical models. It is thus critical to derive an effective model for the charge transfer regime, akin to what has been done in hole doped cuprates \cite{zhang1988effective}.

There is a notable distinction between bilayer nickelates and cuprates. The undoped Ni state is in the spin-one $3d^{8}$ configuration with two electrons in the two $e_g$ orbitals, $(d_{x^2-y^2},d_{z^2})$, strongly coupled together by a Hund coupling $J_H$. Thus doped hole in the oxygen can at best screen the spin-one moment down to spin-half, contrasting the Zhang-Rice singlet in cuprate. In this work, we demonstrate that the doped hole enters the  in-plane oxygen orbital and forms a net spin-half moment together with the Ni spin-one moment at moderate pressure.
The hole state state is dubbed as the \textit{Zhang-Rice spin-half}.
Then by taking the Zhang-Rice spin-half as the primary state of the single-hole $d^{8}L$ state and keeping the spin-triplet parent state of the undoped $d^{8}$ state, the type-II t-J model \cite{zhang2020type} is shown to be the minimal model also in the charge-transfer regime of the doped bilayer nickelates. Besides, we reveal that the oxygen $p_z$ orbital living between the two layers can only mediate a very small inter-layer hopping $t_\perp$ due to symmetry. From the density matrix renormalization group (DMRG) simulation, we show that the pairing gradually decreases with $t_{\perp}$ but remains robust. If we further increase across an optimal value, the doped hole shifts from the intra-layer $p_x,p_y$ oxygen orbital to the inter-layer $p_z$ oxygen and the pairing is suppressed. We finally point out that the trilayer  La$_4$Ni$_3$O$_{10}$ is also described by a trilayer version of type II t-J model in the charge transfer regime.

{\it Charge transfer model.}--- We consider a model on bilayer square lattice (See Fig.~\ref{fig:1}), which includes
 three $p$-orbitals ($2p_x,2p_y,2p_z$) of oxygen as well as the two $E_g$ $d$-orbitals ($3d_{x^2-y^2},3d_{z^2}$) of nickel.   
In the hole picture, we first focus on one-unit cell with two Ni atoms and nine O atoms, the Hamiltonian is,
\begin{eqnarray}
H&=& H_{dp}+\sum_{i,l}
\left[U_1 n_{1;i;l;\uparrow}^d n^d_{1;i;l;\downarrow}
+U_2 n_{2;i;l;\uparrow}^d n_{2;i;l;\downarrow}^d
\right. \nonumber
\\
&&+ 
U^\prime n^d_{1;i;l} n^d_{2;i;l}
-2J_H 
(\vec{s}^d_{1;i;l}\cdot \vec{s}^d_{2;i;l}+\frac{1}{4}n^d_{1;i;l}n^d_{2;i;l})
, \nonumber\\
&&\left.+\sum_{a}
\epsilon_{d,a}n_{a;i,l}^{d}\right]
+\sum_{i',l,\alpha}
\epsilon_{p}n_{\alpha;i';l}^{p}
+\sum_{i}\epsilon_{p;z}n^p_{z;i}, 
\label{eq:local_Ham}
\end{eqnarray}
and
\begin{eqnarray}
    H_{dp}&=& \sum_{i,l,a,\sigma}
    \left[2 t_{dp;a} 
    d^\dagger_{a;i;l;\sigma} p_{a;i;l;\sigma} 
    + \mathrm{H.c.}\nonumber
\right] \\
&+&
\sum_{i,\sigma}
 \left[t_{dp;z} (d^\dagger_{2;i;t;\sigma}
-d^\dagger_{2;i;b;\sigma})p_{z;i;\sigma}+\mathrm{H.c.}\right] \nonumber
\end{eqnarray}
with
\begin{eqnarray*}
p_{1;i;l;\sigma}=\frac{1}{2}\left[
p_{x;i+\frac{\hat{x}}{2};l;\sigma}
-p_{x;i-\frac{\hat{x}}{2};l;\sigma}
-p_{y;i+\frac{\hat{y}}{2};l;\sigma}
+p_{y;i-\frac{\hat{y}}{2};l;\sigma}
\right],
\\
p_{2;i;l;\sigma}=\frac{1}{2}\left[
p_{x;i+\frac{\hat{x}}{2};l;\sigma}
-p_{x;i-\frac{\hat{x}}{2};l;\sigma}
+p_{y;i+\frac{\hat{y}}{2};l;\sigma}
-p_{y;i-\frac{\hat{y}}{2};l;\sigma}
\right],
\end{eqnarray*}
where $a=1,2$ labels $d_{x^2-y^2}$ and $d_{z^2}$ orbital of Ni and $l=t,b$ labels the top and bottom layer. 
$\sigma=\uparrow,\downarrow$ labels the spin.
Here $d^\dagger_{a;i;l;\sigma}$  creates a hole on the nickel site $i$ relative to the $3d^{10}$ state. $p^\dagger_{a;i;l;\sigma}$ creates a hole in a superposition of occupying the four oxygen atom sites $i'$ around the Ni site $i$.  The $p_x,p_y$ orbitals are regrouped as $p_1,p_2$ orbitals living on the Ni site according to the $D_{4h}$ symmetry. One can check that $p_{1},p_{2}$ have the same lattice symmetry as the $d_1, d_2$ orbitals. In supplemental material (SM) Sec.\textcolor{Red}{I}, we provide the details on the symmetry analysis.
 $U$($U'$) is the intra-(inter-) orbital onsite repulsion, and $J_{H}>0$ is the Hund coupling.
$t_{dp}$ is the nearest-neighbor hopping between $d-p$ orbitals. 
There are three different hopping channels, $(d_{1;l},p_{1;l})$, $(d_{2;l},p_{2;l})$, and $(d_{2;t}-d_{2;b},p_{z})$ classified by $D_{4h}$ point group, respectively. The relative size of the onsite energies $\epsilon_{d}$ and $\epsilon_{p}$ decides whether the system is in the charge transfer or Mott Hubbard regime. We consider the case that the undoped parent Ni state is in $3d^{8}$ with two holes in the two $e_g$ orbitals, forming a $S=1$ localized moment. Then under further hole doping, additional holes enter the oxygen $p$ orbitals.  We note that the orbital $p_{1}$ and $p_2,p_z$ have different eigenvalues under the $C_4$ rotation and can not hybridize. So the doped hole enters either $p_1$ or $p_2,p_z$ depending on energetics.

\begin{figure}[t]
    \centering
\includegraphics[width=.95\linewidth]{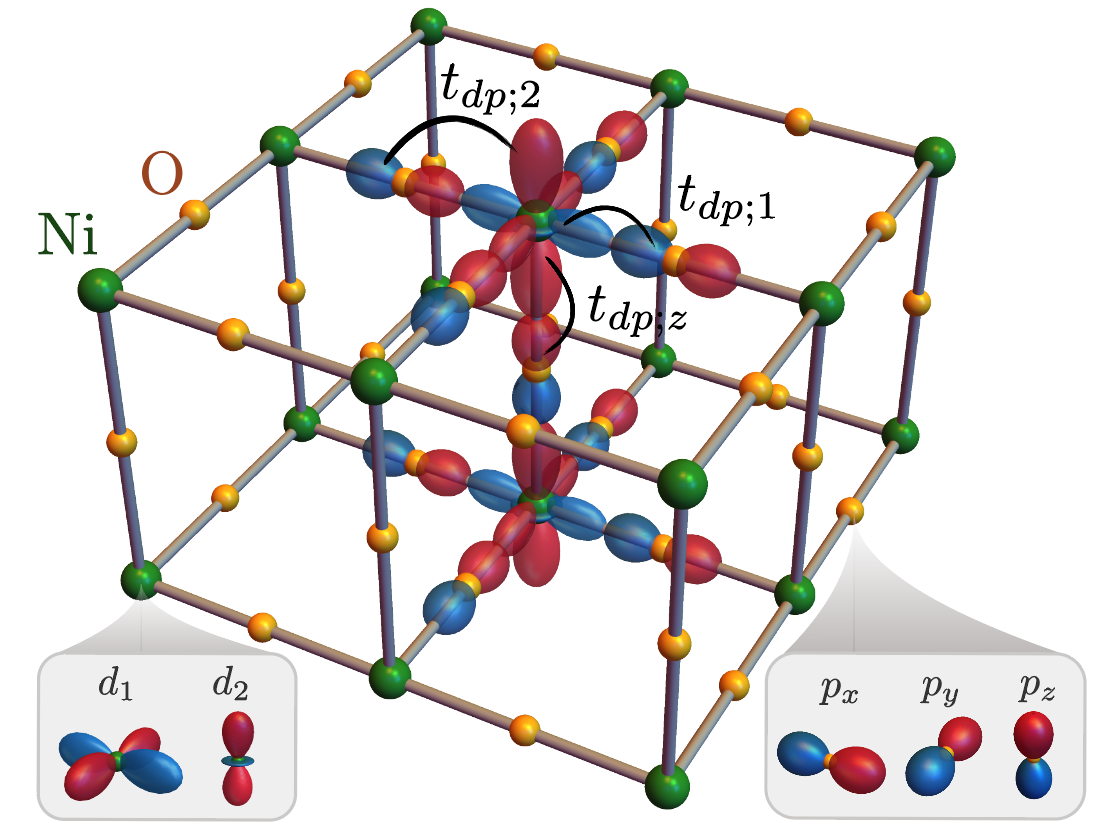}  
    \caption{\textbf{The  lattice structure and $p,d$ orbitals of the bilayer Nickelates La$_3$Ni$_2$O$_7$.} 
    The green (yellow) sphere denotes Ni (O) atom, respectively.
   At each Ni atom, there are two $d$ orbitals, $d_{1},d_{2}$ (Left inset). 
   Around each Ni atom at site $i$, the $p_x,p_y$ orbitals from the four oxygen atoms around the Ni site in the same plane form two orbitals $p_{i;1},p_{i;2}$ living on the same site $i$. Meanwhile, there is a $p_z$ orbital from the oxygen at the center of the $\hat{z}$ bond.  At moderate pressure, the doped hole enters the Wannier orbital, $p_{1;i}=\frac{1}{2}[
p_{x;i+\hat{x}/{2}}
-p_{x;i-\hat{x}/2}
-p_{y;i+\hat{y}/2}
+p_{y;i-\hat{y}/2}
]$ which has the same symmetry as the $d_1$ orbital. The hole in this $p_{i;1}$ orbital couples to the $S=1$ moment of the Ni $3d^8$ state and forms a \textit{Zhang-Rice spin-half} state. }
    \label{fig:1}
\end{figure}
{\it Zhang-Rice Spin-half.}--- In the strong coupling limit, $t_{dp}\ll U,U'$, the hole enters the oxygen $p$ orbitals and interact with the localized spin-one moments from the Ni$^{2+}$ state through a Kondo coupling, 
\begin{eqnarray}
    H_K
    &=&\sum_{i,l,a}
    J_{K;a}\big[(p_{a;i;l}^{\dagger}
    \vec{\sigma}
    p_{a;i;l}) \cdot 
\vec{s}^{d}_{a;i;l}
    -\frac{1}{2}n^{p}_{a;i;l}
\big]\label{eq:Ham_local_spin}
\\
    &+&\sum_{i,l}
J_{K;z}
    [(p_{z;i}^{\dagger}
    \vec{\sigma}
    p_{z;i}) \cdot  \vec{s}^{d}_{2;i;l}
    -\frac{1}{2}n^{p}_{z;i}]\nonumber\\
    &-&
\sum_{i,l}2J_{H}
[\vec{s}^{d}_{1;i;l}\cdot \vec{s}^{d}_{2;i;l}
    +\frac{1}{4}
    ]
    ,\nonumber
\end{eqnarray}
where $\vec s^d_{a;i;l}$ is the spin 1/2 operator from the $d_a$ orbital of the Ni at site $i$ of layer $l=t,b$. We have
\begin{eqnarray}
J_{K;a}&=&\frac{4t^2_{dp;a}}{U_1+U^\prime
+J_H
-\Delta_{p;a}}+\frac{4t^2_{dp;a}}{\Delta_{p;a}+J_H-U^\prime},
\nonumber\\
J_{K;z}&=&\frac{t^2_{dp;z}}{U_2+U^\prime
+J_H
-\Delta_{p;2}^{z}}+\frac{t^2_{dp;z}}{\Delta_{p;2}^{z}
+J_H
-U^\prime}. \label{eq:Kondo}
\end{eqnarray}
with $\Delta_{p;a}=\epsilon_{p}-\epsilon_{d;a}$, $\Delta_{p;a}^{z}=\epsilon_{p;z}-\epsilon_{d;a}$.
Here, we assume $ U'-J_{H}
\ll \Delta_{p,a},\Delta_{p,a}^{z} \ll U_{a}+U'$ so the system is in the charge transfer regime. For illustration, we adopt the hopping parameters calculated at 29.5GPa by DFT, $t_{dp;1}=1.56$eV, $t_{dp;2}=-0.75$eV, $t_{dp;z}=1.63$eV \cite{wu2023charge}.
We assume the similar interaction strength, $U_a=10$eV, $U'=6$eV, $J_{H}=2$eV and as in the cuprates \cite{Ogata_2008} and use $\Delta_{p;a}=\Delta_{p;a}^{z}=9$eV.
Within those parameter sets, we found that $J_{K;z}\simeq J_{K;1}/4$ with $J_{K;1}= 3.03$eV and $J_{K;z}= 0.83$eV.

\begin{figure}[tb]
    \centering
\includegraphics[width=1.0\linewidth]{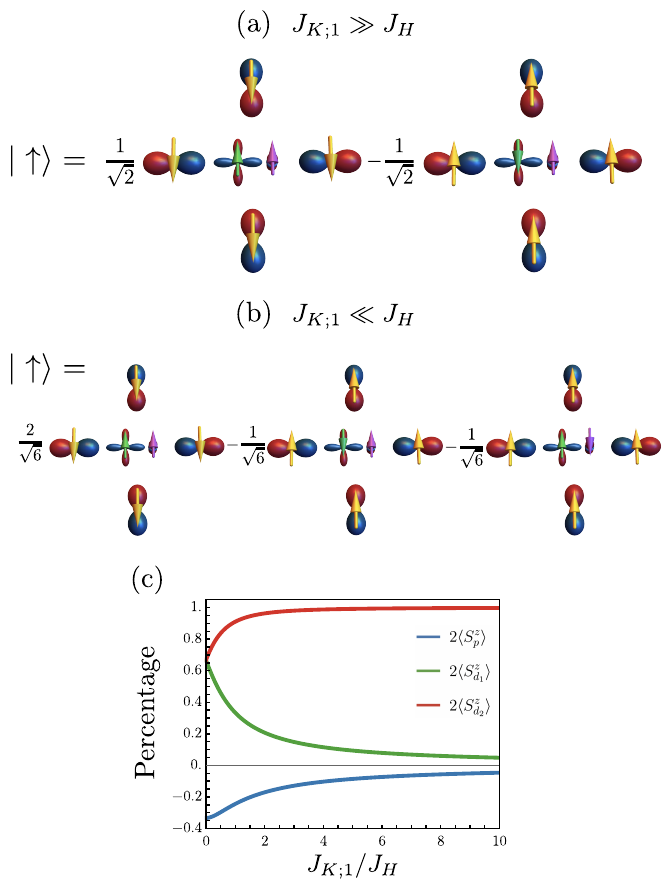}
    \caption{\textbf{
    The wave function of Zhang-Rice spin-1/2 states of the bilayer nickelates}. 
    The local $d_{1}$, $d_{2}$ orbitals of Ni atoms surrounded by the four in-plane $p$ orbitals. The yellow, green, and magenta arrow is for the spin of the $p_1$, $d_1$, and $d_2$ orbital, respectively. 
    (a) In the $J_{K;1}\gg J_{H}$ limit, the Zhang-Rice spin-1/2 state is a simple product state of the Zhang-Rice singlet made by $d_{1},p_{1}$ and the decoupled $d_{2}$ orbital. (b) In the $J_{K;1}\ll J_{H}$ limit, two $d$ orbital forming a spin triplet forms a spin $1/2$ along with $p_1$ orbitals.
    Here, for simplicity, we illustrate only the $\ket{{\uparrow}}$, since $\ket{{\downarrow}}$ is just a its time-reversal partner.
    (c) The three orbital contributions in $S^z_{\mathrm{tot}}=\frac{1}{2}$ of $\ket{\uparrow}$ state. As $J_{K;1}/J_{H}\rightarrow +\infty$, the spin-half state is dominated by the $d_2$ orbital. Generically the state is a combination of all three orbitals: $p_1$, $d_1$, and $d_2$. 
    }
    \label{fig:2}
\end{figure}

We should view $p_1, p_2$ as Wannier orbitals centered on Ni atom. The doped hole is favored to enter one of them or the $p_z$ orbital by minimizing the Kondo coupling $J_{K;1},J_{K;2},J_{K;z}$.
When the hole occupies the in-plane $p_{a}$ orbital, it forms a net $S=\frac{1}{2}$ together with the original spin-one moment from the Ni$^{2+}$.
On the other hand, when the hole occupies the $p_{z}$ orbital, it couples to the spin-one moments of Ni from both layers.  The ground state energies  resulting from the Kondo interactions per unit cell summed over two layers for each case are 
\begin{eqnarray*}
    E_{G}^{a}=
    -\big[J_{H}+\sqrt{J_{H}^2 +
    J_{K;a}J_{H}+
    J_{K;a}^2
    }+J_{K;a}\big], 
\end{eqnarray*}
\begin{eqnarray*}
E_{G}^{z}=-\frac{1}{2}\big[2
J_{H} 
+ \sqrt{4J_{H}^2  + 8 J_{K;z}J_{H} + 9  {J_{K;z}}^2
}+3 J_{K;z}\big].
\end{eqnarray*}
Using the condition $t_{dp;1}\gg t_{dp;2}$, we always have $E_{G}^{1}(J_{K;1})\gg E_{G}^{2}(J_{K;2})$ and the main competition is between the $p_1$ and $p_z$ orbitals. For the estimated parameter of $J_{K;1}\simeq 4J_{K;z}$, we have $E_{G}^{1}(J_{K;1}\simeq 4J_{K;z})<E_{G}^{z}(J_{K;z})$ and the $p_1$ orbital wins. In this regime the doped hole enters the in-plane oxygen orbital $p_1$ and screens the spin-one Ni$^{2+}$ down to spin-half. We dub this hole state as Zhang-Rice spin-half state as analog to the Zhang-Rice singlet in hole doped cuprate. 
This \textit{Zhang-Rice spin-half} consists of the spins from the $p_1$ orbital and the original $d_1$ and $d_2$ orbital of Ni. The wavefunction depends on the parameter $r=J_{K;1}/J_{H}$:
\begin{eqnarray}
     \ket{
\sigma
   }_{l}
\sim  (1+r+\alpha)
\ket{ \overline{\sigma}, \sigma, \sigma
}
-(r+\alpha)
\ket{ \sigma,\overline{\sigma}, \sigma}
-\ket{ \sigma, \sigma, \overline{\sigma}},\!  
\label{eq:zhang_rice}
\end{eqnarray}
with $\sigma=\pm \frac{1}{2}$.
We used $\alpha(r)=[r^2+r+1]^{1/2}$  and here omitted the normalization factor, for simplicity (See SM Sec.\textcolor{Red}{II}). 
Also, $\overline{\sigma}=-\sigma$, and $\ket{\sigma,\sigma_1,\sigma_2}$ denotes the spin of $p_{1;l}, d_{1;l},d_{2;l}$ orbital respectively. 
A few remarks are as follows. 
In the limit of $J_{H}\ll J_{K;1}$, the state is simply the tensor product of the Zhang-Rice singlet from only $d_1,p_1$ orbitals and a decoupled spin-half from the $d_2$ orbitals shown in Fig.~\ref{fig:2}(a). 
However, with a general $J_H$, the three orbitals are highly entangled  and should be considered together.  
For instance, in Fig. \ref{fig:2}(c), we
plot the spin contributions of the net spin-1/2 from all three orbitals.
The bipartite entropy between the $d_1,p_1$ orbitals and the $d_2$ orbital increases with increasing $J_H$, reaching its maximum values in the limit where $J_{H} \gg J_{K;1}$ (See SM), 
where we illustrate the state in Fig.~\ref{fig:2}(b).

The state $|{\sigma}\rangle_l$ is derived from solving the Hamiltonian in a single unit cell, next we should make the states at two different sites $i,j$ orthogonal. We just call this state as $|{\sigma}\rangle_{i;l}$. This can be simply done by replacing the $p_1$ orbital in Eq.(\ref{eq:zhang_rice}) with a wannierized electron operator, $c_{i;l;\sigma}^\dagger
=\mathcal{N'}
\sum_{j}
B(i-j)p_{1;j;l;\sigma}$
where $B(i-j)=\frac{1}{N}\sum_{k}e^{ik\cdot(i-j)}\beta(k)$ and $\beta(k)=[1-\frac{1}{2}(\cos k_x + \cos k_y)]^{-1/2}$ which form orthogonal and complete sets with $\{ c_{i;l;\sigma},c_{j;l';\sigma'}^{\dagger}\}=\delta_{i,j}\delta_{l,l'}\delta_{\sigma,\sigma'}\mathcal{N'}^2$. Here, we use $\mathcal{N}'(r)=[2(r+\alpha)^2+2(r+\alpha)+2]^{1/2}/(r+\alpha+1)$, monotonic increasing function as increasing $r$, ranging $[1.22,1.41]$.
Then, we can turn to the electron picture in the remaining part of the paper, which is more intuitive, through using $c_{i;l;\sigma}$. In this context, we have $n_{i;c}=1-x$, which corresponds to the condition $n_{i;p}=x$ in the hole representation.

{\it Type-II t-J model.}--- Now we can see that the low energy model in the charge transfer regime is still the type II t-J model \cite{zhang2020type}. The undoped Ni site is in the $d^8$ with spin-one, while the single hole $d^8 L$ state corresponds to the Zhang-Rice spin-half state above, which is an analog of  the $d^7$ states in the Mott-Hubbard regime given that the $p_1$ orbital has the same crystal symmetry as the $d_1$ orbital. 
Then, the wannierized orbital $c_{i;l;\sigma}$ can be projected in the $2+3=5$ dimensional Hilbert space. 
We now turn to the electron picture for simplicity. For a general $r=J_{K;1}/J_{H}$, the electron operator of the $p_1$ orbital is written as, 
\begin{eqnarray*}
    c_{i;l;\uparrow}&=& +
    \prod_{j<i}
{(-1)^{n_j}}
    \left[\ket{\downarrow}_{i;l}\bra{-1}_{i;l}
    + \frac{1}{\sqrt{2}}\ket{\uparrow}_{i;l}\bra{0}_{i;l}\right], 
    \label{eq:p_electron_1}
     \\
  c_{i;l;\downarrow}&=& -
    \prod_{j<i}
{(-1)^{n_j}}
\left[\ket{\uparrow}_{i;l}\bra{1}_{i;l}
    + \frac{1}{\sqrt{2}}\ket{\downarrow}_{i;l}\bra{0}_{i;l}\right], 
    \label{eq:p_electron_2}
\end{eqnarray*}
with the Jordan-Wigner string, $\prod_{j<i}(-1)^{n_j}$.
The spin-triplet states of $d^8$ are $\ket{-1}_{l}=d^\dagger_{1;l\downarrow}d^\dagger_{2;l;\downarrow}\ket{G}$, $\ket{0}_{l}=\frac{1}{\sqrt{2}}(d^\dagger_{1;l;\uparrow}d^\dagger_{2;l;\downarrow}+d^\dagger_{1;l;\downarrow}d^\dagger_{2;l;\uparrow})\ket{G}$ and $\ket{1}_{l}=d^\dagger_{1;l;\uparrow}d^\dagger_{2;l;\uparrow}\ket{G}$. Note that $d_{a;l,\sigma}$ is a hole-operator of $d$ orbital and $\ket{G}$ is defined as vacuum as a $3d^{10}$ configuration.

Then, the charge-transferred type II t-J model written in the electron picture is,
\begin{eqnarray}
     H=T_{K}&+&   
     J_\parallel^{ss} \sum_l \sum_{\langle ij \rangle} \vec s_{i;l}\cdot \vec s_{j;l}
     +J^{dd}_\parallel \sum_l \sum_{\langle ij \rangle}\vec S_{i;l}\cdot \vec S_{j;l} \notag \\
&+& J^{sd}_\parallel \sum_l \sum_{\langle ij \rangle} (\vec s_{i;l}\cdot \vec S_{j;l} 
+\vec S_{i;l}\cdot \vec s_{j;l}) \label{eq:type_II_t_J_main}\\ 
&+&
J_\perp^{ss} \sum_{i} \vec s_{i;t}\cdot \vec s_{i;b}
+J^{dd}_\perp \sum_i \vec S_{i;t}\cdot \vec S_{i;b} \notag \\
&+&
J^{sd}_\perp 
 \sum_i (\vec s_{i;t}\cdot \vec S_{i;b}+\vec S_{i;t}\cdot \vec s_{i;b})
 +V\sum_{i}
n_{i;t}n_{i;b}, \notag 
\end{eqnarray}
with 
\begin{eqnarray*}
  T_{K} &=&
    -
    t_{\parallel}
    \sum_{l,\sigma,\langle i ,j \rangle} c_{i;l;\sigma}^{\dagger} 
    c_{j;l;\sigma}
    -t_{\perp}
    \sum_{\sigma,i} 
    c_{i;t;\sigma}^{\dagger} c_{i;b;\sigma}
 +\mathrm{H.c.},
\end{eqnarray*}

The spin operators for the spin-1/2 state are $ \vec s_{i;l}=\frac{1}{2}\sum_{\sigma \sigma'} \ket{\sigma}_{i;l} \vec \sigma_{\sigma \sigma'}\bra{\sigma'}_{i;l}$. 
Meanwhile, the spin operators for the spin-one moment are written as $ \vec S_{i;l}^{d}=\sum_{\alpha,\beta=-1,0,1} \vec T_{\alpha \beta]}\ket{\alpha}_{i;l} \bra{\beta}_{i;l}$. We have $T^x=\frac{1}{\sqrt{2}}\begin{pmatrix} 0 & 1 & 0 \\ 1 & 0 & 1 \\ 0 & 1 & 0 \end{pmatrix}$ and $ T^y=\frac{1}{\sqrt{2}}\begin{pmatrix} 0 & -i & 0 \\ i & 0 & -i \\ 0 & i & 0 \end{pmatrix}$ in the $S^{z}=1,0,-1$ basis. $V$ is the inter-layer repulsive density interaction.
The expressions for $J_{\perp}$, $J_{\parallel}$ in terms of the microscopic parameters are provided with the derivation in SM Sec.\textcolor{Red}{III}. Additionally, we establish the relationships $J^{ss}_{\perp} = \mathcal{C}(r)J^{sd}_{\perp} = \mathcal{C}(r)^2J^{dd}_{\perp}$ and $J^{ss}_{\parallel} = \mathcal{C}(r)\mathcal{A}J^{sd}_{\parallel} = \mathcal{C}(r)^2\mathcal{A}^2 J^{dd}_{\parallel}$, where $\mathcal{A}\simeq 0.7705$. 
Using the parameters mentioned above, we have 
$J_{\parallel}^{dd}=0.097$eV, $J_{\perp}^{dd}$=0.076eV, and $t_{\parallel}$ is in the range of $0.32-0.49$eV. 
$\mathcal{C}(r)$ ranges from $4/3$ at $r\rightarrow 0$ limit and $0$ at $r\rightarrow \infty$ limit. Here $r=J_{K;1}/J_H$.
The dependence of $t_{\parallel}(r)$ and $\mathcal{C}(r)$ are provided in SM.
In the Mott-Hubbard regime, we have 
$J^{ss}_{\perp}=2J^{sd}_{\perp}=4J^{dd}_{\perp}=4t_{\perp}^2/U$, 
and $J^{ss}_{\parallel}=J^{sd}_{\parallel}=0$, $J^{dd}_{\parallel}=t_{\parallel}^2/U$ \cite{yang2023strong}. We stress that the differences in their ratios do not alter the qualitative behavior significantly.

We highlight that even in the charge transfer regime, the type-II t-J model still shows suppressed hopping between the in-plane $p_1$ orbitals of the two layers, with $t_{\perp}\simeq0$, akin to the behavior observed in the Mott-Hubbard regime. This can be validated by a simple symmetry argument that since $p_1$ and $p_z$ orbitals have different symmetries, there is no direct coupling channel between them at the lowest order. A finite $t_{\perp}$ can only be generated through virtual hopping to a $p_z$ orbital at a different site and its value should be small. In the following we will show that a small $t_\perp$ has no significant effect and thus as a good approximation we can set it to be zero.    
\begin{figure}[tb]
    \centering
\includegraphics[width=1.0\linewidth]{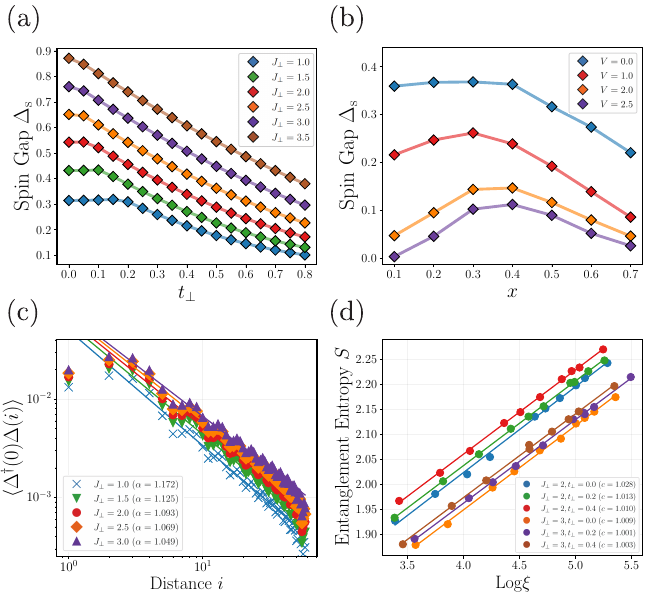}
    \caption{\textbf{DMRG simulation results of the 
    type-II t-J model, Eq.(\ref{eq:type_II_t_J_main}) in the two-leg ladder at $t_{\parallel}=1$, $J^{\parallel}_{ss}=0.1$} Here, we impose $J^{ss}_{\perp} = \frac{4}{3}J^{sd}_{\perp} = \frac{16}{9}J^{dd}_{\perp}$ and $J^{ss}_{\parallel} = \frac{4}{3}\mathcal{A}J^{sd}_{\parallel} = \frac{16}{9}\mathcal{A}^2 J^{dd}_{\parallel}$ with $\mathcal{A}=0.7705$.
    (a) The $t_{\perp}$ dependence of the spin gap, $\Delta_{s}=E(S_z=1)-E(S_z=0)$, of the type-II t-J model at $V=0$ and $x=0.5$ with various $J^\perp_{ss}=1,1.5,2,3,3.5$.
Here, we use $L_x=40$ and $\chi=2400$ for the simulation. The spin gap monotonically decreases as increasing in the $t_{\perp}$ in the large $J_{\perp}$ limit. 
(b) The doping dependence of the spin gap at $t_{\perp}=0.1$, $J_\perp^{ss}=1$ with various $V=0,1,2,2.5$. The superconducting dome is exhibited in the BCS limit, where we increase $V$, and the substantial spin gap remains even in $x=0.5$.  
(c) The pair correlation at $t_{\perp}=0.1$ and $x=0.5$ with various $J_{\perp}^{ss}$ values. It exhibits the power law decaying where the fitted power $\alpha$ is denoted as the solid line. 
(d) The scaling of the entanglement entropy and the correlation length $\xi$ at $x=0.5$ within the infinite DMRG calculation.
Here, we choose $t_\perp=0,0.2,0.4$ and $J_{\perp}=2,3$. 
The extracted central charge $c$ is carried by the relation $S=\frac{c}{6}\log \xi$. 
    }
    \label{fig:3}
\end{figure}

{\it Numerical simulations.}--- We perform the density matrix renormalization group (DMRG) simulations applied to the type-II t-J model described by Eq.(\ref{eq:type_II_t_J_main}) \cite{PhysRevLett.69.2863,10.21468/SciPostPhysLectNotes.5}.  We consider the two-leg ladder configuration ($L_z=2, L_y=1, L_x\rightarrow\infty$), rather than the full bilayer two-dimensional cubic lattice, due to the well-known limitation of the DMRG.
In our simulations, we set $t_{\parallel}=1$ and $J_{\parallel}^{ss}=0.1$. 
In Fig.~\ref{fig:3}, we present the spin gap varying the parameters $J_{\perp}^{ss}$, $t_{\perp}$, and the doping ratio. 
The convergence of the DMRG results under the system size is also checked an provided in the SM Sec.\textcolor{Red}{IV}.
 The presence of a finite spin gap indicates the emergence of the Luther-Emily liquid phase \cite{PhysRevLett.33.589} with power-law inter-layer pairing correlations.  The pairing gap gradually decreases with $t_{\perp}$ because inter-layer pairing frustrates the $t_\perp$ term. But the pairing remains robust at small $t_\perp$. In the bilayer nickelate we believe $t_\perp \ll t$ and hence its effect should be negligible. 
One of the remarkable features of the bilayer type-II t-J model is the doping dependence of the pairing gap. 
As presented in Fig.~\ref{fig:3} (b), the pairing scale shows a dome centered near $x = 0.4-0.5$ in the presence of a repulsion. This is due to the doping induced BCS to BEC crossover\cite{yang2023strong}. Finally, we check the key characteristics of Luther-Emily liquids, such as power-law pair correlation functions and a central charge $c=1$, in Figs.~\ref{fig:3} (c-d). 
In the main figure, we only illustrate the case with $x=0.5$ relevant to the experiments, but the Luther-Emily liquid phases are manifested in the broad range of $x$ (See SM Sec.\textcolor{Red}{IV}).

{\it Discussion.}---  Our theoretical proposal can provide a potential scenario on the dome of the $T_c$ versus pressure in the experiment \cite{li2024pressuredriven}. Increasing pressure should enhance $t_{dp;z}$ and thus the $J_\perp$ term. Initially, $T_c$ increasing with the pressure due to the increase of the $J_\perp$. However, when the pressure is larger than an optimal value $P_*$, the hole prefers to stay in the $p_z$ orbital and then the pairing is suppressed (see SM Sec.\textcolor{Red}{V}). The shift from the in-plane $p_1$ orbital to the $p_z$ orbital is likely the origin of the dome with pressure.

Our analysis can also be generalized  to the trilayer nickelates, La$_4$Ni$_3$O$_{10}$ \cite{zhu2024superconductivity} (see SM Sec.\textcolor{Red}{VI}) and a trilayer version of the type II t-J model is the minimal model. Now the model is very different from the models proposed assuming Mott-Hubbard regime. For the trilayer case both $d_1,d_2$ orbitals are assumed to be at fractional filling and the mobile carriers are argued to be from both orbitals 
\cite{sakakibara2024theoretical,zhang2024prediction,lu2024superconductivity}. In contrast, in the charge transfer picture, both $d_1,d_2$ are still localized which just provide a spin-one moment at each Ni site. Then the doped hole enters the $p_1$ orbital and the final model is still one-orbital like, similar to the bilayer case. We leave to future to analyze this trilayer model.

{\it Conclusion.}---  In summary, we provide analytical and numerical study for bilayer nickelates, La$_3$Ni$_2$O$_7$ within the charge transfer framework. Our primary discovery is the identification of the Zhang-Rice spin-half state as the dominant hole state. We emphasize that the type II t-J model again serves as a minimal model  for both the bilayer and the trilayer nickelate. The physics is distinct from the hole doped cuprates due to the importance of strong inter-layer spin-spin coupling. This leads to an optimal doping as large as $40-50\%$ in the bilayer type II t-J model in contrast to $20\%$ in cuprate.

\textit{Acknowledgement.} 
H.Oh thanks Hui Yang for valuable help on the density matrix renormalization group simulation codes. B.Zhou thanks Xinlong Liu for fruitful discussions on constructing Zhang-Rice spin-half states in the stage of starting the project. 
This work was supported by the National Science Foundation under Grant No. DMR-2237031.

%

\onecolumngrid
\newpage
\clearpage

\setcounter{equation}{0}
\setcounter{figure}{0}
\setcounter{table}{0}
\setcounter{page}{1}

\maketitle 
\makeatletter
\renewcommand{\theequation}{S\arabic{equation}}
\renewcommand{\thefigure}{S\arabic{figure}}
\renewcommand{\thetable}{S\arabic{table}}

\begin{center}
\vspace{10pt}
\textbf{\large Supplemental Material for 
``Type II t-J model in charge transfer regime in bilayer La$_3$Ni$_2$O$_7$ and trilayer La$_4$Ni$_3$O$_{10}$''}
\end{center} 
\begin{center} 
{Hanbit Oh$^{\ \textcolor{red}{*}}$, Boran Zhou, and Ya-Hui Zhang$^{\ \textcolor{red}{\dagger}}$
}\\
\emph{William H. Miller III Department of Physics and Astronomy, \\
Johns Hopkins University, Baltimore, Maryland, 21218, USA}
\vspace{5pt}
\end{center}
\tableofcontents
\section{I. Symmetry properties of the Charge transfer Hamiltonian}
The charge transfer Hamiltonian enjoys 
$D_{4h}=D_{4}\otimes \mathcal{I}$ point group symmetry defined at Ni $i$ site. While the two $d$ orbitals, ($d_{1},d_{2}$) are already one-dimensional irreducible representations of $D_{4h}$, $p_{x,y}$ orbitals living at the link of the cubic lattice are not. 
To classify them as an symmetry representation, we first identify the symmetry actions of the $p$ orbitals. For example, under the $D_{4}=\{E,2C_{4},C_{2},2C_{2}',2C_{2}''\}$, one can show that 
\begin{eqnarray*}
    C_{4}: 
    \quad 
    p_{x;i\pm \frac{x}{2};l}\rightarrow p_{y;i\pm \frac{y}{2};l}, 
    \quad
    p_{y;i\pm \frac{y}{2};l}\rightarrow -p_{x;i\mp \frac{x}{2};l},
\end{eqnarray*}
\begin{eqnarray*}
    C_{2}: 
    \quad 
    p_{x;i\pm \frac{x}{2};l}\rightarrow -p_{x;i\mp \frac{x}{2};l}, 
    \quad
    p_{y;i\pm \frac{y}{2};l}\rightarrow -p_{y;i\mp \frac{y}{2};l},
\end{eqnarray*}
\begin{eqnarray*}
    C_{2}': 
    \quad 
    p_{x;i\pm \frac{x}{2};l}\rightarrow p_{x;i\pm \frac{x}{2};l}, 
    \quad
    p_{y;i\pm \frac{y}{2};l}\rightarrow -p_{y;i\mp \frac{y}{2};l},
\end{eqnarray*}
\begin{eqnarray*}
    C_{2}'': 
    \quad 
    p_{x;i\pm \frac{x}{2};l}\rightarrow p_{y;i\pm \frac{y}{2};l}, 
    \quad
    p_{y;i\pm \frac{y}{2};l}\rightarrow p_{x;i\mp \frac{x}{2};l},
\end{eqnarray*}

Solving the eigenvalues of the above symmetry actions, we can find that the following linear combinations, 
\begin{eqnarray}
p_{1;i;l;\sigma} &=&\frac{1}{2}\left[
p_{x;i+\frac{\hat{x}}{2};l;\sigma}
-p_{x;i-\frac{\hat{x}}{2};l;\sigma}
-p_{y;i+\frac{\hat{y}}{2};l;\sigma}
+p_{y;i-\frac{\hat{y}}{2};l;\sigma}
\right],\label{eq:p1}
\\
p_{2;i;l;\sigma} &=&\frac{1}{2}\left[
p_{x;i+\frac{\hat{x}}{2};l;\sigma}
-p_{x;i-\frac{\hat{x}}{2};l;\sigma}
+p_{y;i+\frac{\hat{y}}{2};l;\sigma}
-p_{y;i-\frac{\hat{y}}{2};l;\sigma}\label{eq:p2}
\right],
\end{eqnarray}
become one-dimensional representations with $B_{1g}$, $A_{1g}$ representation. 
This can be checked by the character table in Table \ref{table:s1}. We tabulate the characters of the each orbitals and the irreducible representation.
Here, $g(u)$ denotes the even (odd) under the inversion $\mathcal{I}$.

The symmetry allowed nearest neighbor hopping between $d,p$ orbitals are given as 
\begin{eqnarray*}
    H_{dp} &=& \sum_{l,\sigma}\sum_{\langle a;i,\alpha;i'\rangle}\left[ 
    t_{dp}^{a;i,\alpha;i'} d^\dagger_{a;i;l;\sigma} p_{\alpha;i';l;\sigma} 
    + \mathrm{H.c.}\nonumber
\right] +
\sum_{\sigma}
\sum_i
 \left[t_{dp;z} (d^\dagger_{2;i;t;\sigma}
-d^\dagger_{2;i;b;\sigma})p_{z;i;\sigma}+\mathrm{H.c.}\right] \nonumber
\end{eqnarray*}
where $i$($i'$) are sites of nickel (oxygen) atoms. The phase factors should be defined accordingly on the hopping terms, 
\begin{eqnarray*}
    t_{dp}^{1;i,\alpha;i'} &=&t_{dp;1}
    \sum_{s=\{\pm1\}}(-1)^{s}
[\delta_{\alpha,x}
\delta_{i',i+s\frac{\hat{x}}{2}}
-\delta_{\alpha,y}
\delta_{i',i+s\frac{\hat{y}}{2}}
] \\
    t_{dp}^{2;i,\alpha;i'} &=&t_{dp;2}
    \sum_{s=\{\pm1\}}(-1)^{s}
[\delta_{\alpha,x}
\delta_{i',i+s\frac{\hat{x}}{2}}
+\delta_{\alpha,y}
\delta_{i',i+s\frac{\hat{y}}{2}}
]. 
\end{eqnarray*}
Substituting the Eqs.(\ref{eq:p1}-\ref{eq:p2}) into $H_{dp}$ simply reduces to,
\begin{eqnarray}
    H_{dp}&=& \sum_{i,l,a,\sigma}
    \left[2 t_{dp;a} 
    d^\dagger_{a;i;l;\sigma} p_{a;i;l;\sigma} 
    + \mathrm{H.c.}\nonumber
\right] +
\sum_{i,\sigma}
 \left[t_{dp;z} (d^\dagger_{2;i;t;\sigma}
-d^\dagger_{2;i;b;\sigma})p_{z;i;\sigma}+\mathrm{H.c.}\right] \nonumber
\end{eqnarray}
which is introduced in the main-text.

  \begin{table}[h]
\centering
\renewcommand{\arraystretch}{1.35}
\begin{tabular}{c|c|cccccc}
\hline \hline
Orbitals &  R &$E$&$2C_{4}(z)$& $C_2$&$2C_{2}'$ &$2C_{2}''$ & $\mathcal{I}$ \\ \hline
$d_{1;l},p_{1;l}$& $B_{1g}$& 1	&-1	&1&1&-1&1\\
$d_{2;l},p_{2;l}$& $A_{1g}$& 1	&1	&1&1&1&1\\
$d_{2;t}-d_{2;b},p_{z}$&$A_{1u}$&1	&1	&1&1&1&-1
\\
\hline
\hline
\end{tabular}
\caption{\textbf{The irreducible representation (R) and the character table of $D_{4h}$ point group.}
 }
\label{table:s1}
\end{table}

\section{II. Zhang-Rice spin-half state with general $J_{K},J_{H}$}
\label{sec1}
In this section, we provide the energy analysis of possible $d^{8}L$ states with general values of $J_{K},J_{H}$. 
In the strong coupling limit $J_{K},J_{H}\gg 1$, we can consider the local Hamiltonian defined at each site, neglecting the hopping term,
\begin{eqnarray}
 H
    &=&\sum_{l; a} 2J_{K;a}[\vec{s}^{d}_{l;a}\cdot \vec{s}^{p}_{l;a}
    -\frac{1}{4}n^{p}_{l;a}
    ]+
 \sum_{l}2J_{K;z}
    [\vec{s}^{d}_{l;2}\cdot \vec{s}^{p}_{z}
    -\frac{1}{4}n^{p}_{z}]
    - \sum_{l}
    2J_{H}
    [\vec{s}^{d}_{l;1}\cdot \vec{s}^{d}_{l;2}
    +\frac{1}{4}
    ],\label{eq:Ham_local_spin2}
    \end{eqnarray}
where $J_{K;a}, J_{K;z}$ is the Kondo coupling derived in main text, Eq.(\textcolor{Red}{3}). 
Based on the parameters listed in Ref.\cite{wu2023charge}, we assume $t_{dp;1}\simeq t_{dp;z}>t_{dp;2}$ and $\Delta_{p;1}\simeq \Delta_{p;2}^{z}\simeq \Delta_{p;2}$, leading to the relation $J_{K;1}\approx 4J_{K;z}>J_{K;2}$.
There are three possible states depending on the $p$ orbital occupancy : $(i)$  $n_{l;1}^{p}=1$, $(ii)$ $n_{l;2}^{p}=1$, and $(iii)$ $n_{l;z}^{p}=1$.

\begin{figure}[h]
    \centering
    \includegraphics[scale=.7]{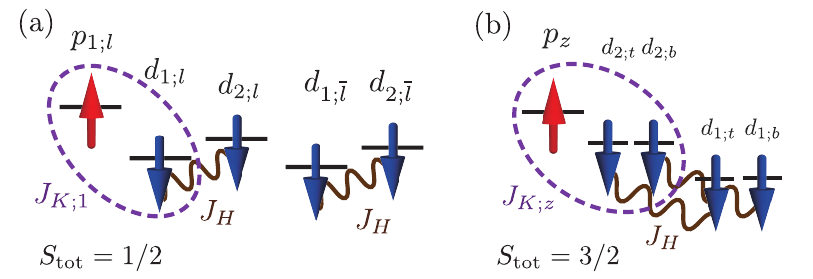}
     \caption{\textbf{Possible Zhang-Rice states of $d^{8}L$ electron configurations of the bilayer nickelates}. 
    (a) The spin $1/2$ and (b) spin $3/2$ states are hybridized by intra-layer $(p_{1},p_2)$ orbital or inter-layer $p_{z}$ orbital of oxygen atoms. 
Note that here all of the states are drawn in the hole description picture.
    }
    \label{fig:s0}
\end{figure}

\begin{figure}[h]
    \centering
    \includegraphics[scale=.7]{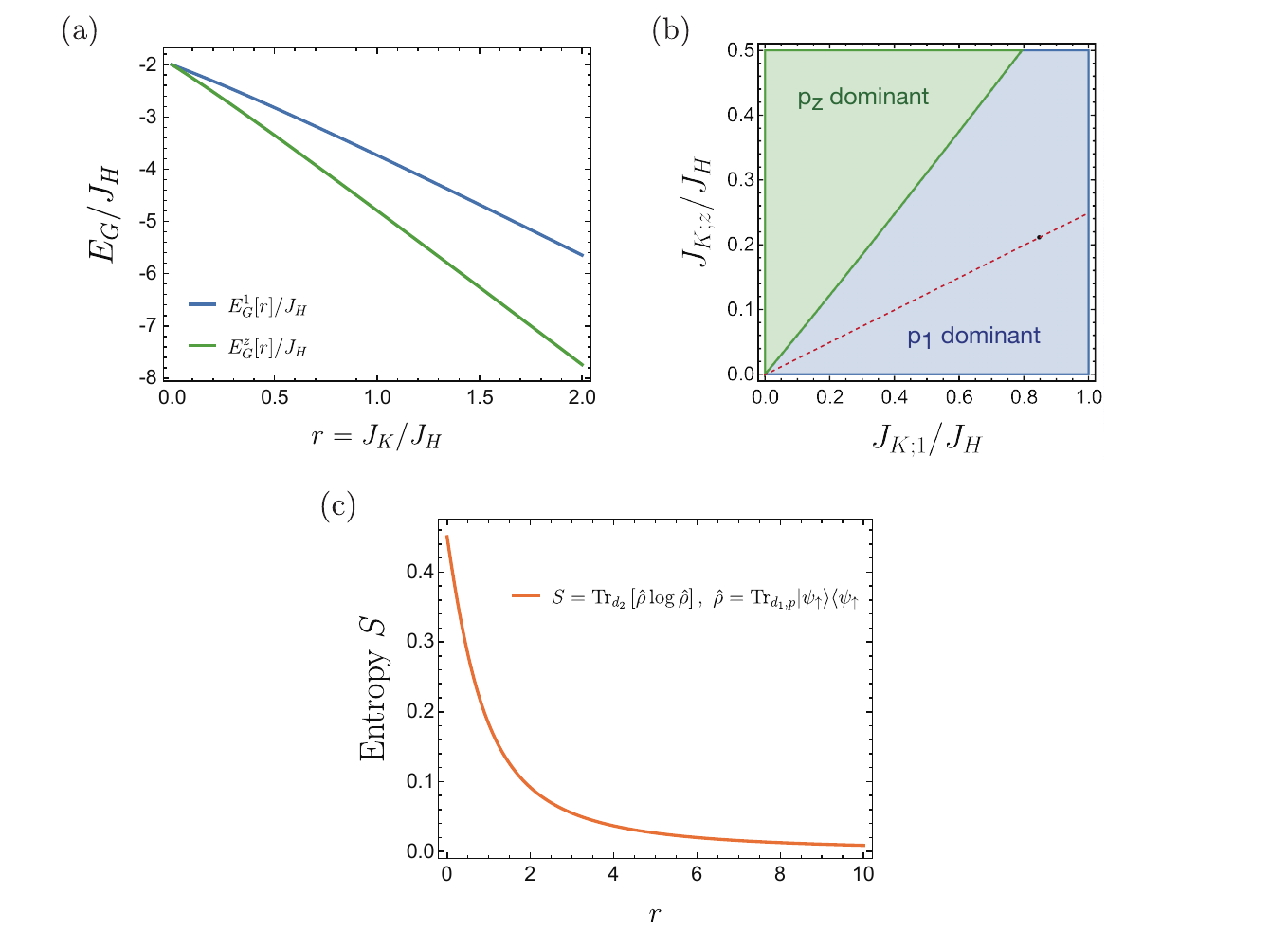}
    \caption{\textbf{(a) The $J_{K}/J_H=r$ dependence of $E_{G}^{1}(r)$ and $E_{G}^{z}(r)$.}
    \textbf{(b)Comparison of the ground states energy in $(J_{K;1},J_{K;z})$ plane.} 
    The blue (green) region indicates the phase space where $p_1$ ($p_z$) orbitals energetically favored, so an additonal holes enter to $p_1$ ($p_z$) orbitals.
    Using the estimation $J_{
K;1}\sim 4 J_{K;z}$ (red dashed line), we conclude that the Zhang-Rice spin-1/2 state with $p_1$ hole is more stable.
\textbf{(c) The bipartite entanglement entropy of $\ket{\uparrow}$, Eqs.(\ref{eq:ZR_up}-\ref{eq:ZR_down}), as a function of $r$.}
Here, the entanglement entropy is evaluated by $S=\mathrm{Tr}_{d_{1}}(\hat{\rho} \log \hat{\rho})$, with the reduced density matrix, $\hat{\rho}=\mathrm{Tr}_{d_1,p}[\ket{\uparrow}\bra{\uparrow}]$. As $r\rightarrow +\infty$ limit, Zhang-Rice spin half state is just a tensor product of the singlet formed by the ($d_1,p_{1}$)
and spin 1/2 of $d_2$ orbital, and the entanglement between two subsystems goes to zero. However, in other value of $r$, the $d_2$ orbital does not be decoupled from $d_1,p_1$ orbitals, forming the Zhang-Rice spin half state together. }
    \label{fig:s1}
\end{figure}

\begin{itemize}
    \item $n_{l;1}^{p}=1$ or $n_{l;2}^{p}=1$ case: 
The local Hamiltonian is reduced into three-site Hamiltonian written in terms of $(p_{a;l},d_{1;l},d_{2;l})$,
\begin{eqnarray*}
 H
    &=&2J_{K;a}[\vec{s}^{d}_{l;a}\cdot \vec{s}^{p}_{l;a}
    -\frac{1}{4}
    ]
    - \sum_{l}2J_{H}
    [\vec{s}^{d}_{l;1}\cdot \vec{s}^{d}_{l;2}
    +\frac{1}{4}
    ]. 
    \end{eqnarray*}
    The ground states of the above Hamiltonian is the spin $1/2$ state whose wave function is given by,  
    \begin{eqnarray}
    \ket{
    { \uparrow}
   }_l
    &= &
 \frac{1}{\sqrt{2[(r+\alpha)^2+(r+\alpha+1)]}}[  (1+r+\alpha)
| \downarrow,\uparrow, \uparrow\rangle
-(r+\alpha)
| \uparrow, \downarrow,\uparrow\rangle
-|\uparrow,\uparrow ,\downarrow\rangle],
\label{eq:ZR_up}
\\
    \ket{{ \downarrow}}_l
    &=&
\frac{-1}{\sqrt{2[(r+\alpha)^2+(r+\alpha+1)]}}[   (1+r+\alpha)
|\uparrow,\downarrow, \downarrow\rangle
-(r+\alpha)
| \downarrow, \uparrow,\downarrow\rangle
-|\downarrow,\uparrow, \downarrow  \rangle].
\label{eq:ZR_down}
\end{eqnarray}
where $\ket{s_0,s_{1},s_{2}}$ implies the spin of the $p_{a},d_{1},d_{2}$ orbital respectively. 
Here, we use $r=J_{K;a}/J_H$ and $\alpha (r)=\sqrt{r^2+r+1}$.
These states are called as Zhang-Rice spin-half state, and the schematic illustrations are provided in Fig.~\ref{fig:s0}. 
The wave function is obviously entangled by the three ($p_1, d_1,d_2$) orbitals in the most range of $r$, except $r=+\infty$. In this specific point $r=+\infty$, the state becomes the tensor product of the Zhang-Rice singlet only made by ($p_1, d_1$) and decoupled $d_2$ orbitals. The subsystem entanglement entropy is obtained in Fig.~\ref{fig:s1}(b), manifesting its values is finite due to the entanglement of ($p_1, d_1$) and  $d_2$. 
The associated energy is obtained as,
\begin{eqnarray}
    E_{G}=-[1+r+\sqrt{r^2+r+1}]J_{H}, \label{e:Energy_a}
\end{eqnarray}
introducing the dimensionless parameter, $r=J_{K;a}/J_{H}$. 

 \item $n_{z}^{p}=1$ case: 
 The local Hamiltonian is reduced into five-site Hamiltonian of $(p_{z},d_{1;t},d_{1;b},d_{2;t},d_{2;b})$,
 \begin{eqnarray*}
 H
    &=&
 \sum_{l}2J_{K;z}
    [\vec{s}^{d}_{l;2}\cdot \vec{s}^{p}_{z}
    -\frac{1}{4}]
    - \sum_{l}2J_{H}
    [\vec{s}^{d}_{l;1}\cdot \vec{s}^{d}_{l;2}
    +\frac{1}{4}
    ]. 
    \end{eqnarray*}
The wave function $\ket{\psi}$ of the ground states of the above Hamiltonian have four-fold degeneracy and are very complicated.  
However, the important thing to note is the total spin of the state is $S_{\mathrm{tot}}= \frac{3}{2}$, i.e. $S_{tot}^2\ket{\psi}=\frac{15}{4}\ket{\psi}$, and $S_{tot}^{z}\ket{\psi}=\pm \frac{1}{2},\pm \frac{3}{2}\ket{\psi}$,  for any values of positive $J_{K;z}$ and $J_{H}$.
The associated ground state energy is obtained as 
\begin{eqnarray}
    E_{G}^{z}=\frac{1}{2}[-2 - 3 r - \sqrt{4 + 8 r + 9 r^2}]J_{H},
    \label{e:Energy_z}
\end{eqnarray}
with a dimensionless parameter, $r=J_{K;z}/J_{H}$,
\end{itemize}

The above-hybridized states are illustrated in Fig. \ref{fig:s0}. 
We now compare those three energies, Eqs.(\ref{e:Energy_a}-\ref{e:Energy_z}), using the fact $J_{K;1}\approx 4J_{K;z}>J_{K;2}$.
Obviously, $E_{G}^{1}(J_{K;1})<E_{G}^{2}(J_{K;2})$ is established.
Next, we compare the energy between $E_{G}^{1}$ and $E_{G}^{z}$ in Fig. \ref{fig:s1} showing that $E_{G}^{1}(J_{K;1})<E_{G}^{z}(J_{K;z}\simeq 4J_{K;1})$.
Hence, we can conclude that the Zhang-Rice spin-1/2 state, especially with $p_{1}$ hole is the most stable state.

\section{III. Derivation of type-II t-J model from the charge transfer Hamiltonian}\label{sec2}
In this section, we derive the type-II t-J model starting from the charge transfer Hamiltonian, Eq. (\textcolor{Red}{1}). 
In particular, we show all the spin-exchange couplings of the type-II t-J model, in terms of the original Hamiltonian. 
The spin-exchange interaction of two nickel atoms can be derived by the fourth-order perturbation in the strong coupling limit.  
We divide the charge-transfer Hamiltonian into the kinetic part, $V=H_{dp}$, and remaining part, $H_{0}=H-V$.

\subsection{A. Derivation of $J_{\perp}$}
\begin{figure}[h]
    \centering
\includegraphics[width=1\linewidth]{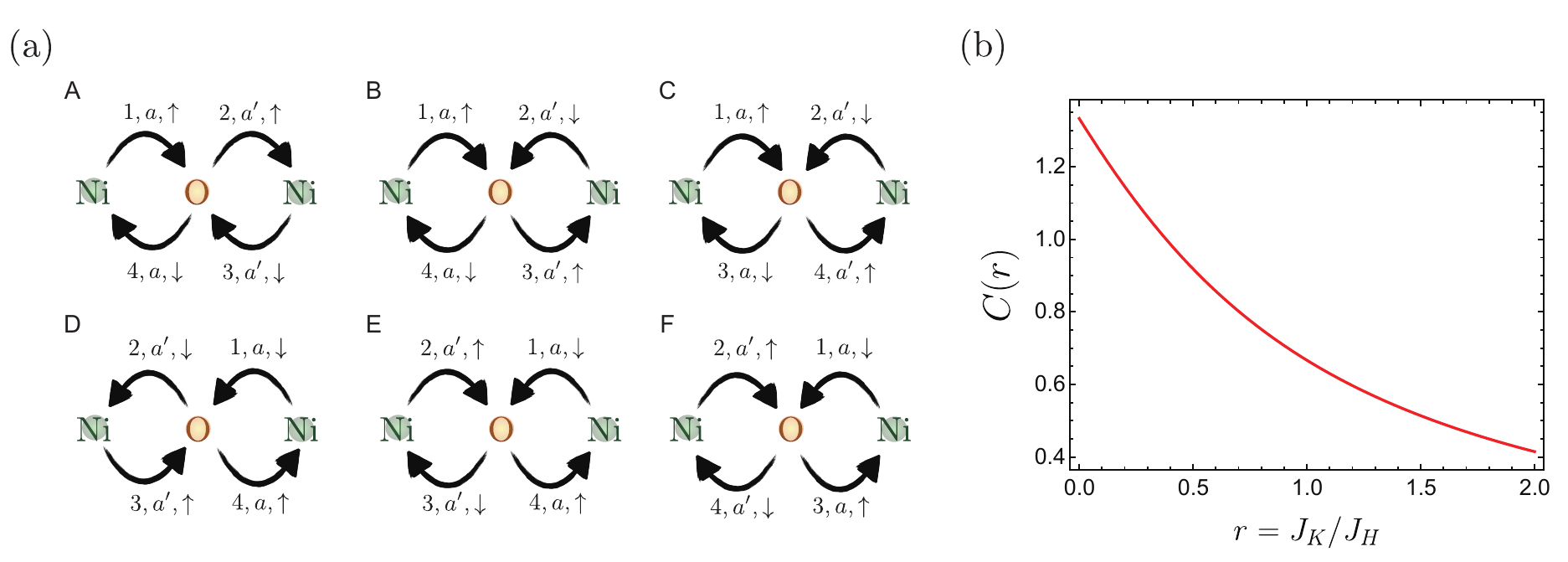}
    \caption{
\textbf{(a) All possible processes contributing to the spin exchange interaction upon the fourth-order perturbation.} The green (yellow) sphere represents an oxygen (nickel) atom. 
In $(i,a,\sigma)$, $i$ denotes the order of the process, $a=1,2$ is for labeling $d_a$ orbitals and $\sigma$ is for the spin. 
These processes specifically contribute to the $s^{-}_{L} s^{+}_{R}$, $s^{-}_{L} S^{+}_{R}$ or $S^{-}_{L} S^{+}_{R}$ where $L(R)$ denotes the left (right) nickel atom. \textbf{(b) The $J_{K}/J_H=r$ dependence of $\mathcal{C}(r)$ in Eqs.(\ref{eq:Jsd_perp}-\ref{eq:Jss_perp})}. In particular, $\mathcal{C}(r)$ becomes 
$\frac{4}{3}$ in the $r\rightarrow 0 $ limit. 
}

\label{fig:s2}
\end{figure}

First, we show how the $J^{\perp}$ terms are derived, exemplified by $J^{\perp}_{sd}$, and the generalization into $J^{\perp}_{ss},J^{\perp}_{dd}$ is straightforward. Due to the SU(2) symmetry of the Hamiltonian, it is enough to evaluate the coefficient of one of the components of $\frac{1}{2}(s^{-}_{t}S^{+}_{b})=\frac{1}{\sqrt{2}} \ket{s_t=-\frac{1}{2},s_b=0}\bra{s_t=\frac{1}{2},s_b=-1}+\cdots $ piece, not all $\vec{s}_{t;i}\cdot \vec{S}_{b;i}$. 
There are six different processes (A)-(F) contributing the $s^{-}_{t}S^{+}_{b}$, as illustrated in Fig.\ref{fig:s2} (a),
\begin{eqnarray}
      \bra{s_t=-\frac{1}{2},s_b=0}
 VRVR VRV\ket{s_t =\frac{1}{2}, s_{b}=-1}&=& 
(A)+(B)+(C)+(D)+(E)+(F),
\end{eqnarray}
where $R=Q/(E_{0}-H_{0})$ is the projection onto the subspace of all exited states. 
Here, $Q=1-P$ is a complementary operator of the projection operator of the ground state of $H_{0}$ where the doubly occupied states are forbidden.

With some tedious calculations, we find each component is given by 
\begin{eqnarray*}
(A)&=&
\frac{t_{dp,z}}{(\Delta_{p;2}^{z}+\overline{J}_{H})}
    \bra{-\frac{1}{2}}_{t}
\bra{0}_{b} VRVRV
p_{z;\uparrow}^{\dagger}
[
\frac{(1+r+\alpha)}{\mathcal{N}}
\ket{s_{d_1}=\frac{1}{2},s_{c}=-\frac{1}{2}
}_{t}
-\frac{1}{\mathcal{N}}
\ket{s_{d_1}=-\frac{1}{2},s_{c}=\frac{1}{2}
}_{t}]
\ket{-1}_{b}\nonumber\\
 &=& \frac{(t_{dp,z})^2}{(\Delta_{p;2}^{z}+\overline{J}_{H})(U+2J_{H})}
 \bra{-\frac{1}{2}}_{t}
\bra{0}_{b}  VRV
d_{2;b;\uparrow}^{\dagger}
[
\frac{(1+r+\alpha)}{\mathcal{N}}
\ket{s_{d_1}=\frac{1}{2},s_{c}=-\frac{1}{2}
}_{t}
-
\frac{1}{\mathcal{N}}
\ket{s_{d_1}=-\frac{1}{2},s_{c}=\frac{1}{2}
}]_{t}\ket{-1}_{b}\nonumber\\
 &=& \frac{-(t_{dp,z})^3}{(\Delta_{p;2}^{z}+\overline{J}_{H})^2 (U+2J_{H})}
 \bra{-\frac{1}{2}}_{t}
\bra{0}_{b}  V p_{z;\downarrow}^{\dagger}
[
(1+r+\alpha)
\ket{s_{d_1}=\frac{1}{2},s_{c}=-\frac{1}{2}
}_{t}
-
\ket{s_{d_1}=-\frac{1}{2},s_{c}=\frac{1}{2}
}_{t}]
\frac{\ket{0}_{b}}{\sqrt{2}}\nonumber\\
&=&\frac{(t_{dp,z})^4}{(\Delta_{p;2}^{z}+\overline{J}_{H})^2 (U+2J_{H})}\frac{\sqrt{2}(1+r+\alpha)}{\mathcal{N}^2}=(D),
\end{eqnarray*}
and
\begin{eqnarray*}
(B)&=&
 \frac{(t_{dp,z})^2}{2(\Delta_{p;2}^{z}+\overline{J}_{H})^2}
 \bra{-\frac{1}{2}}_{t}
\bra{0}_{b}  VRV
p_{z;\downarrow}^{\dagger}
p_{z;\uparrow}^{\dagger}
[\frac{(1+r+\alpha)}{\mathcal{N}}
\ket{s_{d_1}=\frac{1}{2},s_{c}=-\frac{1}{2}
}_{t}
-
\frac{1}{\mathcal{N}}
\ket{s_{d_1}=-\frac{1}{2},s_{c}=\frac{1}{2}
}_{t}]
\ket{s_{d_1}=-\frac{1}{2}}_{b}\nonumber\\
 &=& \frac{-(t_{dp,z})^3}{2(\Delta_{p;2}^{z}+\overline{J}_{H})^3}
 \bra{-\frac{1}{2}}_{t}
\bra{0}_{b}  V p_{z;\downarrow}^{\dagger}
[\frac{(1+r+\alpha)}{\mathcal{N}}
\ket{s_{d_1}=\frac{1}{2},s_{c}=-\frac{1}{2}
}_{t}
-
\frac{1}{\mathcal{N}}
\ket{s_{d_1}=-\frac{1}{2},s_{c}=\frac{1}{2}
}_{t}]\ket{s_{p_{z}}=-\frac{1}{2}}\frac{\ket{0}_{b}}{\sqrt{2}}\nonumber\\
&=&\frac{(t_{dp,z})^4}{2(\Delta_{p;2}^{z}+\overline{J}_{H})^3}
\frac{\sqrt{2}(1+r+\alpha)}{\mathcal{N}^2}=(C)=(E)=(F),
\end{eqnarray*}
with $\alpha =\sqrt{r^2+r+1}$ and $\mathcal{N}=\sqrt{2[(r+\alpha)^2+(r+\alpha+1)]}$.
By putting everything together, $J_{sd}^{\perp}$ is obtained as, 
\begin{eqnarray}
    J_{sd}^{\perp}&=&
    \mathcal{C}(r)
      \frac{t_{dp,z}^4}{(\Delta_{p;2}^{z}+\overline{J}_{H})^2}  \left[ 
    \frac{1}{U_2+2J_{H}}
+\frac{1}{\Delta_{p;2}^{z}+\overline{J}_{H}}
    \right],
    \label{eq:Jsd_perp}
\end{eqnarray}
with $ \mathcal{C}(r)=4(1+r+\alpha)/\mathcal{N}^2$, whose $r$ dependence is illustrated in Fig. \ref{fig:s2} (b). 
Similarly, we can show that 
\begin{eqnarray}
    J_{dd}^{\perp}&=&
      \frac{t_{dp,z}^4}{(\Delta_{p;2}^{z}+\overline{J}_{H})^2}  \left[ 
    \frac{1}{U_2+2J_{H}}
+\frac{1}{\Delta_{p;2}^{z}+\overline{J}_{H}}
    \right],   \label{eq:Jdd_perp}\\
J_{ss}^{\perp}&=&
\mathcal{C}(r)^2
      \frac{t_{dp,z}^4}{(\Delta_{p;2}^{z}+\overline{J}_{H})^2}  \left[ 
    \frac{1}{U_2+2J_{H}}
+\frac{1}{\Delta_{p;2}^{z}+\overline{J}_{H}}
    \right], \label{eq:Jss_perp}
\end{eqnarray}
where we have $J_{ss}^{\perp}=
\mathcal{C}(r) J_{sd}^{\perp}
=\mathcal{C}(r)^2 J_{dd}^{\perp}
$. Here we use 
$\overline{J}_{H}=J_{H}-U'$,  $\Delta_{p;a}=\epsilon_{p}-\epsilon_{d;a}$, and $\Delta_{p;a}^{z}=\epsilon_{p;z}-\epsilon_{d;a}$.

\subsection{B. Derivation of $J_{\parallel}$}
Next, we show the derivation of  $J^{\parallel}$ terms.
There are two significant differences, compared to $J^{\perp}$. 
First, there are two kinds of hopping processes with $t_{dp;1}$ and $t_{dp;2}$. Second, the p orbital of the Zhang-Rice spin-half state can be affected by the hoppings, since the intermediate-occupied oxygen is also shared by those states. 
For example, the hopping process consisting of $p_{x;i+\frac{\hat{x}}{2};\sigma}^{\dagger}$ eliminates the piece of the same component in 
$\phi_{i;\sigma}
=
\sum_{j}
B(i-j)p_{1;j;\sigma}=c^{\dagger}_{i;\sigma}$, where $B(i-j)=\frac{1}{N}\sum_{k}e^{ik\cdot(i-j)}\beta(k)$ and $\beta(k)=[1-\frac{1}{2}(\cos k_x + \cos k_y)]^{-1/2}$. To be specific, consider the following example, 
\begin{eqnarray*}
  p_{x;i+\frac{\hat{x}}{2};\sigma}^{\dagger}
   \phi_{i,\sigma'}^{\dagger}
   &=&
 p_{x;i+\frac{\hat{x}}{2};\sigma}^{\dagger}
 \left[
\sum_{j\in \{i,i+\hat{x}\}^{c} } B(i-j)
p_{1;j,\sigma'} ^{\dagger}
+\sum_{j\in \{i,i+\hat{x}\} } B(i-j)
\overline{p}_{1;j,\sigma'} ^{\dagger}
\right]
=
p_{x;i+\frac{\hat{x}}{2};\sigma}^{\dagger} 
\overline{\phi}_{i,\sigma'}^{\dagger}, 
\end{eqnarray*}
where we define the $\overline{\phi} (\overline{p})$  as the $\phi (p)$, but excluding the $p_{x;i+\hat{x}/2}$ piece,
\begin{eqnarray*}
      \overline{p}_{1;i,\sigma'} ^{\dagger}
&=&
\frac{1}{2}
\left[-p_{x;i-\frac{\hat{x}}{2};l;\sigma}
-p_{y;i+\frac{\hat{y}}{2};l;\sigma}
+p_{y;i-\frac{\hat{y}}{2};l;\sigma}
\right]=   p_{1;i,\sigma'} ^{\dagger}
-
\frac{1}{2}
p_{x;i+\frac{\hat{x}}{2};l;\sigma},
\end{eqnarray*}
\begin{eqnarray*}
\overline{p}_{1;i+\hat{x},\sigma'} ^{\dagger}
&=&
\frac{1}{2}
\left[
p_{x;i+\frac{3\hat{x}}{2};l;\sigma}
-p_{y;i+\hat{x}+\frac{\hat{y}}{2};l;\sigma}
+p_{y;i+\hat{x}-\frac{\hat{y}}{2};l;\sigma}
\right]
=
 p_{1;i,\sigma'} ^{\dagger}
+
\frac{1}{2}
p_{x;i+\frac{\hat{x}}{2};l;\sigma},
\end{eqnarray*}
and
\begin{eqnarray*}    
\overline{\phi}_{i,\sigma'} ^{\dagger}
=
\phi_{i,\sigma'} ^{\dagger}
-[\frac{B(0)-B(1)}{2}]p_{x;i+\frac{x}{2};\sigma}^{\dagger}.
\end{eqnarray*}
Note that $\bra{\overline{\phi}}\phi\rangle \simeq 1-[\frac{B(0)-B(1)}{2}]^2\simeq 0.7705\equiv \mathcal{A}$.
Then, the next step is straightforward to consider all the possible processes depicted in Fig \ref{fig:s2} (a). 
After finishing all the calculations with $\Delta_{p;1}=\Delta_{p;2}$, we obtained that 
\begin{eqnarray}
J_{dd}^{\parallel}
&=&\sum_{a,a'} \frac{t_{dp;a}^2 t_{dp;a'}^2}{(\Delta_{p}+\overline{J}_{H})^2} 
\left[ 
    \frac{1}{U_{a'}+2J_{H}}
+\frac{1}{\Delta_{p}+\overline{J}_{H}}
    \right],
\end{eqnarray}
\begin{eqnarray}
J_{sd}^{\parallel}
&=&
\mathcal{C}(r)
\mathcal{A}\sum_{a,a'} \frac{t_{dp;a}^2 t_{dp;a'}^2}{(\Delta_{p}+\overline{J}_{H})^2} 
\left[ 
    \frac{1}{U_{a'}+2J_{H}}
+\frac{1}{\Delta_{p}+\overline{J}_{H}}
    \right], 
\end{eqnarray}
\begin{eqnarray}
J_{ss}^{\parallel}
&=&(\mathcal{C}(r)\mathcal{A})^2\sum_{a,a'}
      \frac{t_{dp;a}^2 t_{dp;a'}^2}{(\Delta_{p}+\overline{J}_{H})^2} 
\left[ 
    \frac{1}{U_{a'}+2J_{H}}
+\frac{1}{\Delta_{p}+\overline{J}_{H}}
    \right]
\label{eq:Jss_parallel}
\end{eqnarray}
Now, we have $J_{ss}^{\parallel}=
\mathcal{C}(r)\mathcal{A} J_{sd}^{\parallel}
=\mathcal{C}(r)^2 \mathcal{A}^2 J_{dd}^{\parallel}
$. We stress this condition is generally satisfied without imposing $\Delta_{p;1}=\Delta_{p;2}$.

\subsection{C. Derivation of $t_{\parallel}$}
\begin{figure}
    \centering
    \includegraphics[width=0.78\linewidth]{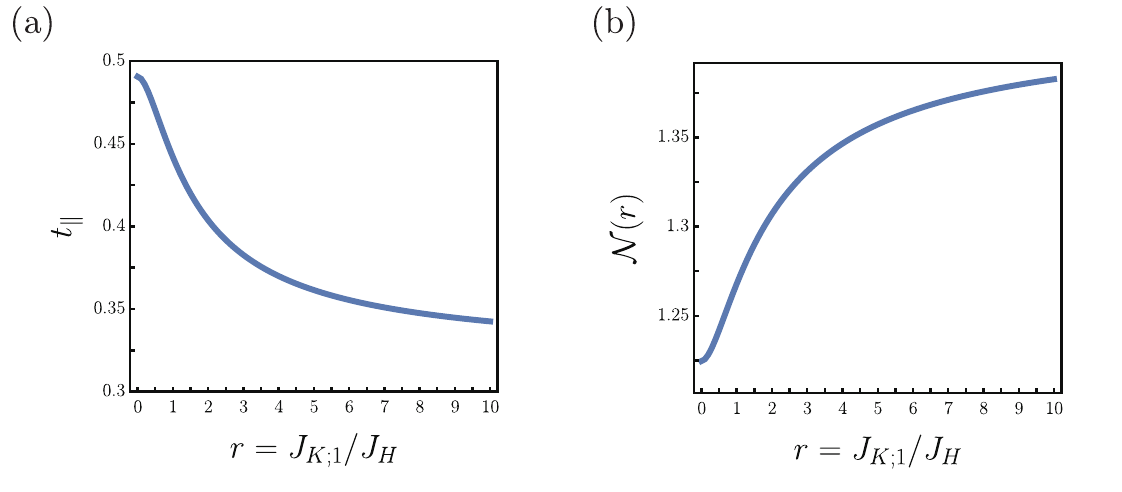}
    \caption{
\textbf{The $r$ dependence of $t_{\parallel}(r)$ and $\mathcal{N}(r)$ estimated by the parameters mentioned in the main-text.}
Note that $t_{\parallel}$ is plotted in the eV unit, and $\mathcal{N}$ is a dimensionless quantity. 
    }
    \label{fig:s9}
\end{figure}

From the second order perturbation, we can derive the in-plane hopping term with and without spin dependence $\sum_{l,\sigma,i,j}-t_{\parallel;i,j} c^\dagger_{i;l;\sigma}c_{j;l,\sigma}$. The main contribution can be classified as the effective O-hopping and spin exchange between the Ni and O holes, similar as the case in cuprate \cite{zhang1988effective}. We have that:
\begin{eqnarray}
    t_{\parallel;i,j}=&\left(\frac{2t_{dp;1}^2}{3(U+U'+J_H-\Delta_{p})}+\frac{J_{K;1}}{12\sqrt{2}}\right)\delta_{ij,\mathrm{NN}}/\mathcal{N}^2(r)-\frac{4J_{K;1}\lambda}{3N}\sum_{\mathbf{k}}\beta^{-1}_{\mathbf{k}}e^{-\mathrm{i}\mathbf{k}\cdot(\mathbf{r_i-r_j})}/\mathcal{N}^4(r),
\end{eqnarray}
in which $\delta_{ij,\mathrm{NN}}$ means that only nearest neighbor term is non-zero. $\lambda$ is defined as $\sum_{\mathbf{k}}\beta^{-1}_{\mathbf{k}}\simeq0.96$. 
Here, we used $\mathcal{N}(r)=[2(r+\alpha)^2+2(r+\alpha)+2]^{1/2}/(r+\alpha+1)$, a function as increasing $r$, which depends on $r=J_{K;1}/J_H$.
The function is monotonic increasing in the range of $[\sqrt{\frac{3}{2}},\sqrt{2}]$. 
$N$ is the number of the site. In our main text, we only keep the nearest neighbor term of $t_{\parallel;i,j}$, and denoted by $t_{\parallel}$.
The $r$ dependnence is illustrated in Fig.\ref{fig:s9}.

\section{IV. More detailed DMRG simulation results}
\begin{figure}
    \centering
    \includegraphics[width=.75\linewidth]{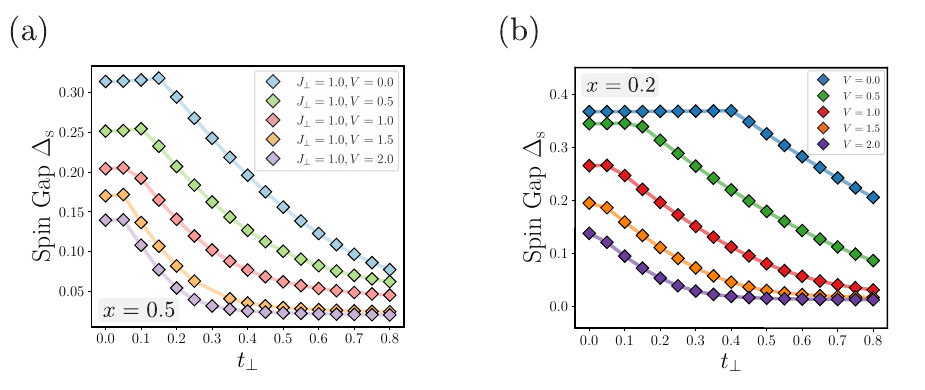}
    \caption{
      \textbf{DMRG simulation results of the type-II t-J model with $L_z=2, L_y=1$ at $t_{\parallel}=1$, $J_{\parallel}^{ss}=0.1$.}
    (a,b) The $t_{\perp}$ dependence of the spin gap with various $V=0,0.5,1,1.5,2$ at  (a) $x=0.5$ and (b) $x=0.2$.
Here, we choose $t_\perp=1$, and use $L_x=40$ and $\chi=2400$ for simulation.
The non-monotonically decreasing region is decreasing as increasing the repulsive interaction where the system goes to the BCS limit.
}
    \label{fig:s4}
\end{figure}

\begin{figure}
    \centering
\includegraphics[width=.8\linewidth]{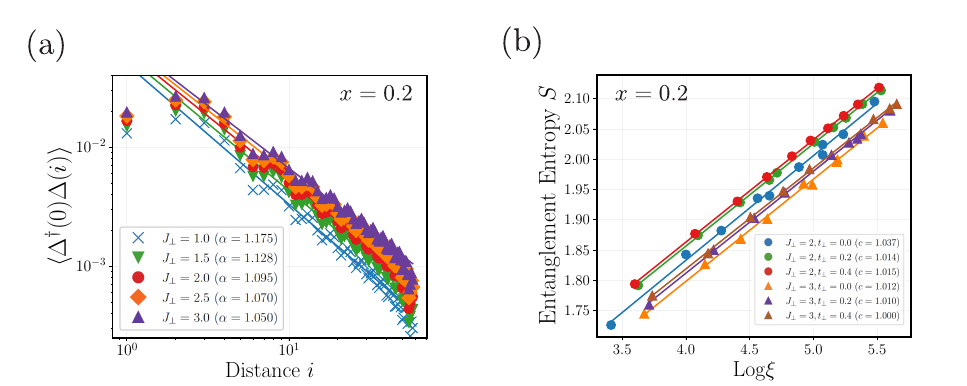}
    \caption{
   \textbf{DMRG simulation results of the type-II t-J model, with $L_z=2, L_y=1$ at $t_{\parallel}=1$, $J_{\parallel}^{ss}=0.1$.} 
The doping is fixed at $x=0.2$ to compare the data at $x=0.5$ provided in the main-text. At $x=0.2$, the results exhibits the evidence of the Luther-Emily liquid phases. 
(a) The pair correlation function at $t_{\perp}=0.1$ and $x=0.2$ shows the powe-law decaying behaviours. 
(b) The entanglement entropy and the correlation length at $x=0.2$. The fitted central charge is nearly $c=1$. 
}
    \label{fig:s5}
\end{figure}

\begin{figure}
    \centering
    \includegraphics[width=.9\linewidth]{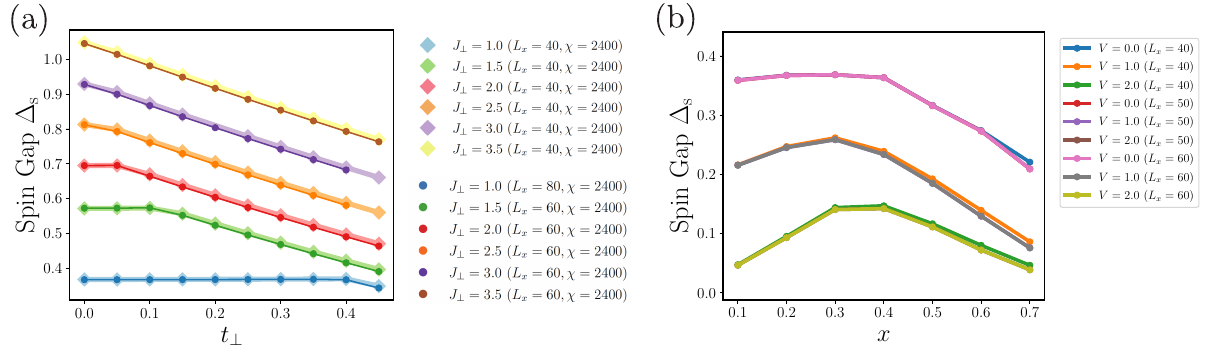}
    \caption{ \textbf{The system size dependence 
    of DMRG simulation results of the type-II t-J model with $L_z=2, L_y=1$ at $t_{\parallel}=1$, $J_{\parallel}^{ss}=0.1$ }
    (a) $t_\perp$ dependence of the spin gap with various $J_{\perp}$ at $x=0.2$. (b) Doping $x$ dependence of the spin gap with various $V$ at $J_{\perp}=1$ ($\chi=2400$).  
    }
    \label{fig:s8}
\end{figure}

In this section, we provide additional DMRG calculation results, which haven't been shown in the main text. Here, we still use $t_{\parallel}=1$ and $J_{\parallel}^{ss}=0.1$.
In Fig.~\ref{fig:s4}, we have shown the $t_\perp$ dependence with various interlayer repulsive interaction strength $V$ to demonstrate, that the non-monotonic decreasing region diminishes and eventually disappears increasing $V$. 
In Fig.~\ref{fig:s5}, we provide some DMRG simulation data at $x=0.2$, for comparing the data provided in the main text at $x=0.5$ and showing that the key characteristic behaviors are similar.   
In Fig.~\ref{fig:s8}, we provide the convergence check under the system size of the spin-gap results through DMRG calculations.

\newpage
\section{V. The pressure dependence of critical temperature
}
\begin{figure}[t]
    \centering
\includegraphics[width=.85\linewidth]{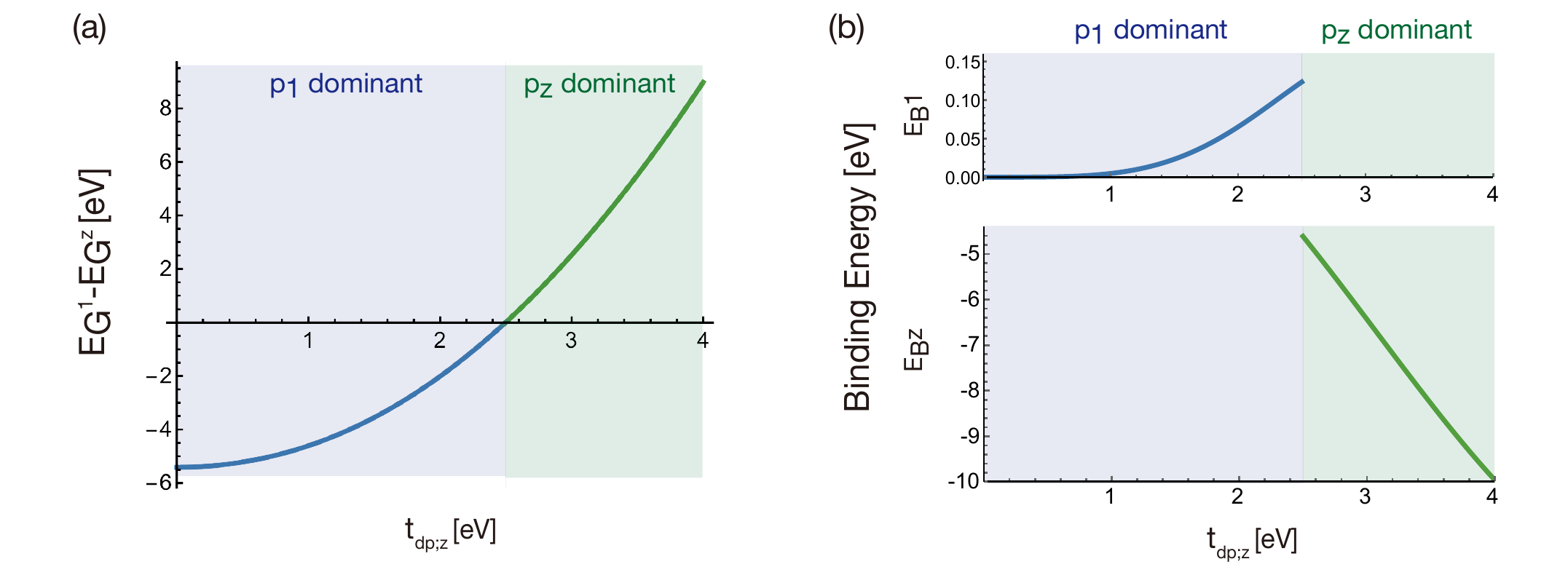}
    \caption{(a) \textbf{$t_{dp;z}$ dependence of $E_G^1-E_G^z$ per unit cell.} There is a transition that hole prefers to stay from $p_1$ to $p_z$ at $t_{dp;z}^{*}=2.50$eV. Here, we used the parameter introduced in the main-text, $t_{dp;1}=1.56$eV, $t_{dp;2}=-0.75$eV, $t_{dp;z}=1.63$eV, $U_a=10$eV, $U'=6$eV, $J_{H}=2$eV, and $\Delta_a=9$ eV. 
    (b) \textbf{$t_{dp;z}$ dependence of the binding energy $E_{B}$ per unit cell.}
    In the $p_1$ dominant regime, the binding energy $E_{B,1}$ is increasing, implying the rise of $T_c$ under the pressure. 
    Meanwhile in the $p_z$ dominant regime, the binding energy $E_{B,z}$ tends to decrease under the pressure. }
    \label{fig:s6}
\end{figure}

In this section, we analyze the binding energy for predicting the critical temperature under changing the pressure. 
Applying pressure effectively increases the hopping between $p_z$ and inplane $d$ orbitals.
Hence, according to our energy estimations, we expect there is a crossover pressure, satisfying $E_{G}^{1}=E_{G}^{z}$, as illustrated in Fig.\ref{fig:s6} (a). 
At $P<P_{*}$, the additional holes are occupied in
$p_1$ orbital, while at $P>P_{*}$, the $p_z$ orbital becomes dominant.

Then, we analyze the binding energy the estimate the critical temperature.
For each regime of $P<P_*$ and $P>P_*$, the binding energy is differently estimated, since the additional hole is occupied by $p_1$ and $p_z$ respectively,
\begin{eqnarray*}
    P<P*: 
    \quad 
   E_{B}^{1}=E[n_{p_1}=1]-(E[n_{p_1}=2]+E[n_{p_1}=0])/2,
   \\
    P>P*: 
    \quad 
   E_{B}^z=E[n_{p_z}=1]-(E[n_{p_z}=2]+E[n_{p_z}=0])/2. 
\end{eqnarray*}
The pressure, (or equivalently $t_{dp;z}$) dependence of the binding energy per unit cell, is illustrated in Fig. \ref{fig:s6} (b). In the regime where $p_1$ dominates, the binding energy exhibits a positive trend, steadily increasing under pressure. This  arises from the stabilization of the two-hole state, occupied by $p_1$ at each layer, which enhanced inter-layer spin coupling by larger $t_{dp;z}$.
Conversely, in the regime where $p_z$ dominates, the binding energy demonstrates a negative trend, decreasing under pressure. 
This stems from the fact that the two hole states occupied by one $p_z$ orbital blocks the hopping between two layers of nickel, potentially leading the inter-layer spin coupling as zero.

\section{VI. Discussion on trilayer La$_4$Ni$_3$O$_{10}$ }
In this section, we expand the theoretical framework of trilayer nickelates La$_4$Ni$_3$O$_{10}$ in the charge-transfer regime. According to the DFT data, the valence of the Ni atom of La$_4$Ni$_3$O$_{10}$ is in the $3d^{7.33}$ configuration (Ni$^{+2.67}$) with the $d_{x^2-y^2}$ orbital is close to one-third occupied $n_1=1/3$ per site, 
while the $d_{z^2}$ orbital is nearly half-filled $n_2\approx 1$ \cite{sakakibara2024theoretical,zhang2024prediction,lu2024superconductivity}.
\begin{figure}[h]
    \centering
\includegraphics[width=0.52\linewidth]{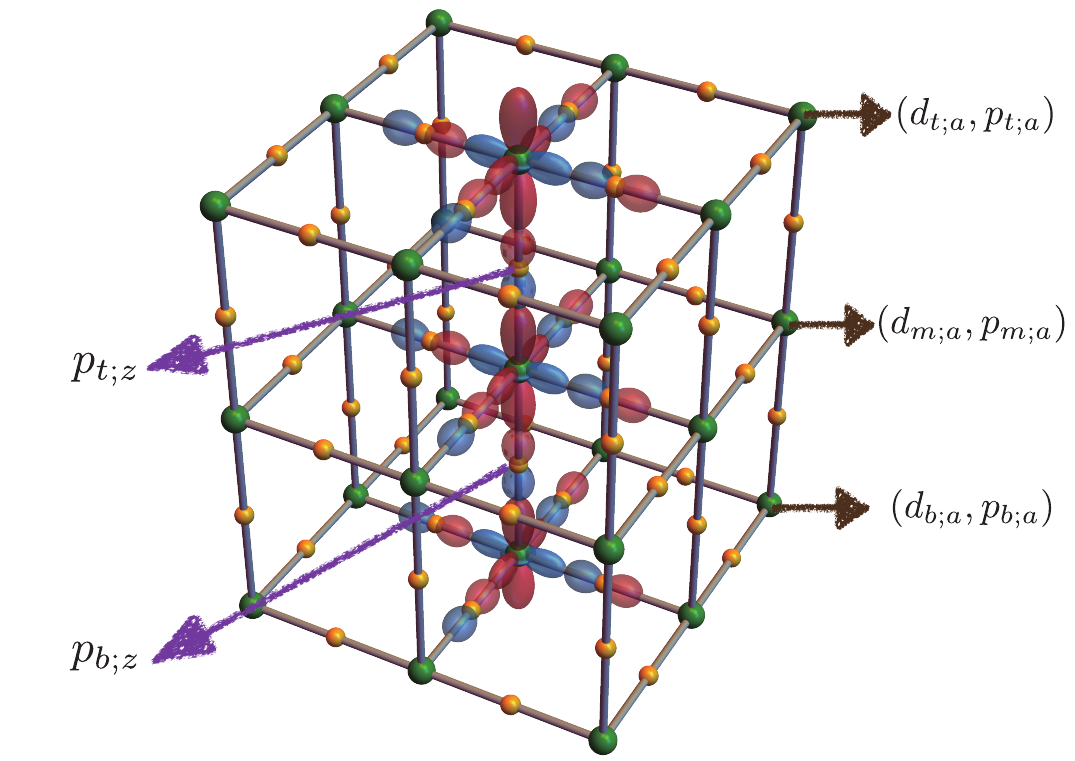}
    \caption{\textbf{The  lattice structure and $p,d$ orbitals of the bilayer Nickelates La$_4$Ni$_3$O$_{10}$.} 
    The green (yellow) sphere denotes Ni (O) atom, respectively.
   At each Ni atom, there are two $d$ orbitals, $d_{1},d_{2}$ at top/middle/bottom layers.
   At each O atom at each layer, either $p_{x}$,$p_{y}$ orbitals reside and can hybridize with the Ni atom at each layer. Meanwhile, the O atoms living in the middle of the three layers have $p_z$ and should be shared by two Ni atoms at adjacent layers. In our effective charge-transfer model, we focus on the 17 atoms consisting of 3 Ni and 14 O atoms. 
    }
    \label{fig:s7}
\end{figure}

\subsection{A. Zhang-Rice spin-half state}
We consider the trilayer cubic lattice with nickel $d_{l;1},d_{l;2}$ orbitals and oxygen $p_{l;1},p_{l;2}$ orbitals defined at each layer $l=t,m,b$. The oxygen $p_{l';z}$ orbitals are living between each layer with  $l'=t,b$ (See Fig.~\ref{fig:s7}). 
Now, in each unit cell, we have to consider the total 17 sites of atoms consisting of three Ni atoms and fourteen O atoms. 
Starting from the local trilayer version of charge-transfer Hamiltonian generalized from Eq.(\textcolor{Red}{1}) in the main text, we again obtain the Kondo Hamiltonian, 
\begin{eqnarray}
 H
    &=&\sum_{l=t,m,b}\sum_{a} 2J_{K;a}[\vec{s}^{d}_{l;a}\cdot \vec{s}^{p}_{l;a}
    -\frac{1}{4}n^{p}_{l;a}
    ]+
 \sum_{l=t,b}2J_{K;z}
    [(\vec{s}^{d}_{l;2}+\vec{s}^{d}_{m;2})\cdot \vec{s}^{p}_{z}
    -\frac{1}{2}n^{p}_{l;z}]
    - \sum_{l=t,m,b}
    2J_{H}
    [\vec{s}^{d}_{l;1}\cdot \vec{s}^{d}_{l;2}
    +\frac{1}{4}
    ],\label{eq:Ham_local_spin2}
    \end{eqnarray}
where $J_{K;a}, J_{K;z}$ is expressed in Eq.(\textcolor{Red}{3}) in the main-text. 
The only difference compared to the bilayer model is that $p_z$ orbital now has a layer index. 
There are still three possible one-hole states depending on the $p$ orbital occupancy : $(i)$  $n_{l;1}^{p}=1$, $(ii)$ $n_{l;2}^{p}=1$, and $(iii)$ $n_{l;z}^{p}=1$. 
For the first two cases ($n_{l;1}^{p}=1$ or $n_{l;2}^{p}=1$) with $l=t,m,b$, the local Hamiltonian is
\begin{eqnarray*}
 H
    &=&2J_{K;a}[\vec{s}^{d}_{l;a}\cdot \vec{s}^{p}_{l;a}
    -\frac{1}{4}
    ]
    - \sum_{l'=t,m,b}2J_{H}
    [\vec{s}^{d}_{l';1}\cdot \vec{s}^{d}_{l';2}
    +\frac{1}{4}
    ],
    \end{eqnarray*}
whose ground state energy is obtained as,
\begin{eqnarray}
    E_{G}=-[2+r+\sqrt{r^2+r+1}]J_{H},
\end{eqnarray}
with a dimensionless parameter, $r=J_{K;a}/J_{H}$. 
For the last case ($n_{l;z}^{p}=1$) with $l=t,b$, the local Hamiltonian is, 
 \begin{eqnarray*}
 H
    &=&
2J_{K;z}
    [(\vec{s}^{d}_{l;2}+\vec{s}^{d}_{m;2})\cdot \vec{s}^{p}_{l;z}
    -\frac{1}{2}]
    - \sum_{l'=t,m,b}2J_{H}
    [\vec{s}^{d}_{l';1}\cdot \vec{s}^{d}_{l';2}
    +\frac{1}{4}
    ],
    \end{eqnarray*}
whose ground state energy is obtained as, 
\begin{eqnarray}
    E_{G}^{z}=\frac{1}{2}[-4 - 3 r - \sqrt{4 + 8 r + 9 r^2}]J_{H},
\end{eqnarray}
with a dimensionless parameter, $r=J_{K;z}/J_{H}$. 
The above results are just a constant shift by $-J_{H}$ from the  Eqs.(\ref{e:Energy_a}-\ref{e:Energy_z}), which leads to the same conclusion. 
Based on the parameters listed in Ref.\cite{luo2024trilayer}, we assume $t_{dp;1}\simeq t_{dp;z}>t_{dp;2}$ and $\Delta_{p;1}\simeq \Delta_{p;2}^{z}\simeq \Delta_{p;2}$, leading to the relation $J_{K;1}\approx 4J_{K;z}>J_{K;2}$.
Using the fact $J_{K;1}\approx 4J_{K;z}>J_{K;2}$, we again have $E_{G}^{1}(J_{K;1})<E_{G}^{2}(J_{K;2})$ and  $E_{G}^{1}(J_{K;1})<E_{G}^{z}(J_{K;z}\simeq 4J_{K;1})$.
Hence, we can conclude that the Zhang-Rice spin-1/2 state, especially with $p_{1}$ hole is the most stable state even for the trilayer nickelates.

\subsection{B. Trilayer layer Type-II t-J model}
Then by taking the Zhang-Rice spin-half as the primary state of the $d^{8}L$ state and keeping a spin-triplet doublon state of the $d^{8}$ state, the minimal model of the hole-doped trilayer bilayer nickelates is the type-II t-J model \cite{zhang2020type,PhysRevB.108.174511}.
The trilayer type-II t-J model Hamiltonian is given by 
\begin{eqnarray}
     H=
H_{K}&+&  \sum_{l=t,m,b} \sum_{\langle i,j \rangle} 
\left[J_{ss}^\parallel 
\vec s_{i;l}\cdot \vec s_{j;l}
+J_{sd}^\parallel(\vec s_{i;l}\cdot \vec S_{j;l} +\cdot \vec S_{i;l}\cdot \vec s_{j;l})
+J_{dd}^\parallel\vec S_{i;l}\cdot \vec S_{j;l}
\right]
\\
&+&
\sum_{l=t,b} \sum_i 
\left[
J_{ss}^\perp  
\vec s_{i;l}\cdot \vec s_{i;m} +J_{sd}^\perp 
 (\vec s_{i;l}\cdot \vec S_{i;m}+\vec S_{i;l}\cdot \vec s_{i;m})
 +J_{dd}^\perp 
 \vec S_{i;l}\cdot \vec S_{i;m} 
\right]
+V\sum_{l=t,b} \sum_{i}n_{i;l}n_{i;m},
\notag
\label{eq:type_II_t_J_tri}
\end{eqnarray}
with
\begin{eqnarray*}
    H_{K} &=&
    -t_{\parallel}
    \sum_{l=t,m,b}
    \sum_{\sigma,\langle i,j \rangle} c_{i;l;\sigma}^{\dagger} 
    c_{j;l;\sigma}
    -t_{\perp}
    \sum_{l=t,b}
    \sum_{\sigma,i} 
    c_{i;l;\sigma}^{\dagger} c_{i;m;\sigma}
 +\mathrm{H.c.},
\end{eqnarray*}
where the coefficients $J_{\parallel},J_{\perp}$ are listed in Eqs.(\ref{eq:Jsd_perp}-\ref{eq:Jss_parallel}).

\end{document}